\documentclass[format=acmsmall, review=false, screen=true]{acmart}

\usepackage{booktabs} 
\usepackage{makecell}
\usepackage{lipsum}
\usepackage[utf8]{inputenc}
\usepackage{array}
\usepackage{wrapfig}
\usepackage{multirow}
\usepackage{tabularx}
\usepackage{pifont}
\usepackage{lineno}
\usepackage{balance}
\usepackage{subcaption}
\usepackage{xcolor,pifont}
\usepackage{caption}
\usepackage{soul}

\usepackage[linesnumbered, ruled,vlined]{algorithm2e}
\RequirePackage{pdflscape}
\usepackage{flushend}
\usepackage{multicol}
\usepackage{placeins}
\usepackage{booktabs}
\usepackage{indentfirst}
\usepackage[figuresright]{rotating}
\usepackage{fancybox}
\usepackage{fontawesome}
\usepackage{threeparttable}
\usepackage{mathtools}
\usepackage{MnSymbol,wasysym}
\usepackage{rotating}
\usepackage{adjustbox}
\usepackage{graphicx}

\usepackage[most]{tcolorbox}
\usepackage{longtable}
\usepackage{setspace}
\usepackage{tcolorbox}
\tcbuselibrary{skins, breakable}

\newtcolorbox[auto counter]{promptbox}[2][]{
  enhanced,
  breakable,
  width=\textwidth,
  colback=gray!10,
  colframe=blue!45!gray,
  boxrule=0.8pt,
  arc=2pt,
  left=8pt,
  right=8pt,
  top=6pt,
  bottom=6pt,
  colbacktitle=blue!45!gray,
  coltitle=white,
  fonttitle=\scriptsize\bfseries,
  title={Prompt \thetcbcounter: #2},  
  label=#1,
  before upper={\scriptsize\setstretch{1.05}\ttfamily}
}

\usepackage{framed}
\definecolor{Gray}{gray}{0.9}
\definecolor{shadecolor}{gray}{0.95}
\usepackage{tikz}
\usetikzlibrary{trees,positioning,shapes,shadows,arrows}
\tikzset{
  basic/.style  = {draw, text width=2cm, drop shadow, font=\sffamily, rectangle},
  root/.style   = {basic, rounded corners=2pt, thin, align=center, fill=white},
  level-2/.style = {basic, rounded corners=6pt, thin,align=center, fill=white, text width=3cm},
  level-3/.style = {basic, thin, align=center, fill=white, text width=1.8cm}
}
\newcommand{\todo}[1]{}
\renewcommand{\todo}[1]{{\color{red} TODO: {#1}}}
\begin{document}
\title{Beyond Functional Correctness: Design Issues in AI IDE-Generated Large-Scale Projects}

\author{Syed Mohammad Kashif}
\orcid{0009-0001-0243-0720}
\affiliation{%
  \institution{School of Computer Science, Wuhan University}
  \city{Wuhan}
  \country{China}
}
\email{smkashif145@gmail.com}

\author{Ruiyin Li}
\orcid{0000-0001-8536-4935}
\affiliation{%
  \institution{School of Computer Science, Wuhan University}
  \city{Wuhan}
  \country{China}
  }
\email{ryli_cs@whu.edu.cn}

\author{Peng Liang}
\orcid{0000-0002-2056-5346}
\affiliation{%
  \institution{School of Computer Science, Wuhan University}
  \city{Wuhan}
  \country{China}
  }
\email{liangp@whu.edu.cn}

\author{Amjed Tahir}
\orcid{0000-0001-9454-1366}
\affiliation{%
  \institution{School of Mathematical and Computational Sciences, Massey University}
  \city{Palmerston North}
  \country{New Zealand}
  }
\email{a.tahir@massey.ac.nz}

\author{Qiong Feng}
\orcid{0000-0003-1667-8062}
\affiliation{%
  \institution{School of Computer Science, Nanjing University of Science and Technology}
  \city{Nanjing}
  \country{China}
}
\email{qiongfeng@njust.edu.cn}

\author{Zengyang Li}
\orcid{0000-0002-7258-993X}
\affiliation{%
  \institution{School of Computer Science, Central China Normal University}
  \city{Wuhan}
  \country{China}
}
\email{zengyangli@ccnu.edu.cn}

\author{Mojtaba Shahin}
\orcid{0000-0002-9081-1354}
\affiliation{%
  \institution{School of Computing Technologies, RMIT University}
  \city{Melbourne }
  \country{Australia}
}
\email{mojtaba.shahin@rmit.edu.au}


\setcopyright{acmlicensed}
\copyrightyear{2026}
\acmYear{2026}
\acmDOI{XXXXXXX.XXXXXXX}

\acmJournal{TOSEM}
\acmVolume{0}
\acmNumber{0}
\acmArticle{0}
\acmMonth{0}

\renewcommand{\shortauthors}{Kashif et al.}

\begin{abstract}
New generation of AI coding tools, including AI-powered IDEs (AI IDEs) equipped with agentic capabilities, can generate code within the context of the project. These AI IDEs are increasingly perceived as capable of producing project-level code at scale. However, there is limited empirical evidence on the extent to which they can generate large-scale software systems and what design issues such systems may exhibit. To address this gap, we conducted an empirical study to explore the capability of Cursor in generating large-scale projects and to evaluate the design quality of projects generated by Cursor as an AI IDE. First, we propose a Feature-Driven Human-In-The-Loop (FD-HITL) framework that systematically guides project generation from curated project descriptions. We generated 10 projects using Cursor with the FD-HITL framework across three application domains (mobile, Web, and utility) and multiple technologies (e.g., React, Spring Boot, Django, and React Native). We assessed the functional correctness of these projects through manual evaluation, obtaining an average functional correctness score of 91\%. Next, we analyzed the generated projects using two static analysis tools, CodeScene and SonarQube, to detect design issues. We identified 1,305 design issues categorized into 9 categories by CodeScene and 3,193 issues in 11 categories by SonarQube. Our findings show that (1) when used with the FD-HITL framework, Cursor can generate functional large-scale projects averaging 16,965 lines of code (LoC) and 114 files; (2) the generated projects nevertheless contain design issues that may pose long-term maintainability and evolvability risks, requiring careful review by experienced developers; (3) the most prevalent issues include \textit{Code Duplication}, high \textit{Code Complexity}, \textit{Large Methods}, \textit{Framework Best-Practice Violations}, \textit{Exception-Handling Issues} and \textit{Accessibility Issues}; and (4) these design issues violate design principles such as the \textit{Single Responsibility Principle (SRP)}, \textit{Separation of Concerns (SoC)}, and \textit{Don't Repeat Yourself (DRY)}. Finally, we contributed DIinAGP, a curated dataset of 10 project descriptions and the corresponding Cursor-generated projects available at \url{https://github.com/Kashifraz/DIinAGP}.
\end{abstract}





\maketitle

\begin{sloppypar}

\section{Introduction} \label{introduction}
The integration of Artificial Intelligence (AI) techniques, especially those employing Large Language Models (LLMs), has transformed the software development paradigm. LLMs have been widely adopted to develop tools that assist software engineers in their development tasks. Developers are increasingly relying on AI code generation tools to write code \cite{daniotti2026using, hou2024llm4se, yellin2023premature}. The most recent Stack Overflow Developer Survey (2025) also indicates significant adoption of AI code generation tools in practice \cite{SO-survey2025}. Previous LLM-assisted AI code generation tools, such as GitHub Copilot \cite{Copilot} and CodeWhisperer \cite{CodeWhisperer}, and general-purpose LLMs (e.g., ChatGPT \cite{ChatGPT} and Gemini \cite{Gemini}), can generate code snippets from natural language descriptions with minimal human intervention \cite{dakhel2023github, tian2023chatgpt}. These code snippets have been reported to improve developers’ productivity and reduce development time and effort \cite{weber2024significant, UsabilityOfAIProgramming}. However, previous AI code generation tools can only generate code snippets \cite{tao2025retrieval, kashif2025developers} and lack the capability to generate project-level code directly in the development environment \cite{tao2025retrieval, kim2025codeassistbench}. This is partly because these tools are not directly integrated into the development environment or lack sufficient development context \cite{li2024enhancing, eibl2025exploring}. 

Recently, a new generation of code generation tools has been introduced, including AI-powered IDEs (e.g., Cursor \cite{Cursor}, Claude Code \cite{Claude}, and OpenAI Codex \cite{Codex}) \cite{li2025rise} and autonomous systems for software development (e.g., MetaGPT \cite{hong2023metagpt} and AgileCoder \cite{nguyen2025agilecoder}). These tools have demonstrated the ability to perform development tasks at the project level within the project context, which can assist developers in building more complex software systems \cite{li2025rise}. In this study, we primarily focus on AI IDEs due to their agentic capabilities, such as interacting with the project environment (e.g., reading codebases, running tools, planning changes) \cite{li2025rise}. Specifically, we focus on Cursor for this study, as this AI IDE has been used in prior research \cite{he2025speed, kumar2025evaluating}. In addition, a recent large-scale empirical study also found that Cursor-generated pull requests (PRs) are predominantly feature implementation PRs \cite{li2025rise}. Developers often operate AI IDEs using \textit{Vibe Coding}, which is a new programming approach in which they generate code without fully understanding the output, prioritizing speed and experimentation, which can affect code quality \cite{fawzy2025vibe}. Despite the potential of these AI IDEs for project-level code generation, two research gaps remain. \textbf{(1)} The existing literature has not explored or evaluated the potential of these AI IDEs for large-scale project generation. Most studies that examine end-to-end full-project generation focus on generating simple (in terms of architectural components and technology stacks) and small projects (in terms of lines of code (LoC))  \cite{hong2023metagpt, nguyen2025agilecoder, rasheed2024codepori}. The descriptions for generating projects are also short and simple, and do not capture the complexity of typical industrial-level projects \cite{hong2023metagpt, rasheed2024codepori}. \textbf{(2)} Existing studies have extensively evaluated AI-generated code, exploring their quality issues \cite{tambon2025bugs, ouyang2024empirical} and security vulnerabilities \cite{yu2024large, pearce2022asleep}. He \textit{et al.} also found that the adoption of Cursor leads to an increase in project-level development velocity but also increases code complexity in GitHub projects \cite{he2025speed}. However, there is a lack of empirical design quality evaluation of project-level code. Since project-level code is generated at scale within the project development context, it is more likely that high-level design issues will arise.
To this end, our study \textbf{aims} to explore the use of Cursor for generating large-scale projects and to evaluate the design quality of the Cursor-generated projects. 
We define a \textbf{large-scale project} as a software system that contains at least 8K lines of code (LoC), follows specific architectural styles with multiple architectural components, and is built with a technology stack (e.g., MERN, LAMP) similar to industrial software systems. We address the following two Research Questions (RQs):

\begin{itemize}
    \item \textbf{RQ1:} To what extent can Cursor generate large-scale projects?
    \item \textbf{RQ2:} What design issues are found in large-scale projects generated by Cursor?
\end{itemize}

To address our RQs, we designed a feature-driven, Human-In-The-Loop (FD-HITL) framework to systematically generate large-scale projects using Cursor. First, we curated 10 project descriptions to generate the projects using Cursor (specifically Cursor Pro). We employed two static analysis tools, CodeScene \cite{CodeScene} and SonarQube \cite{SonarQube}, to detect design issues in the generated projects. We also manually checked the identified design issues and removed 1,612 false positives from the issues identified by SonarQube. We identified 1,305 design issues by CodeScene and 3,193 by SonarQube. We employed quantitative and qualitative techniques to analyze the identified design issues and obtained 9 categories of design issues by CodeScene and 11 by SonarQube. We also reported 133 overlapping design issues between CodeScene and SonarQube. 

Our \textbf{findings} show that: \textbf{(1)} Cursor can generate functional large-scale projects when using our proposed FD-HITL framework. \textbf{(2)} Cursor-generated projects contain significant design issues that may pose long-term maintainability and evolvability challenges and require close supervision by experienced developers with rigorous code reviews. \textbf{(3)} The top design issues we found are \textit{Code Duplication}, \textit{Large Methods}, \textit{Complex Methods}, \textit{Framework Best-Practice Violations}, \textit{Exception-Handling Issues}, \textit{Design Principle Violations}, and \textit{Accessibility Issues}, among others.  

The main \textbf{contributions} of this work are: \textbf{(1)} a proposed Feature-Driven Human-In-The-Loop (FD-HITL) framework for systematically generating large-scale projects using a popular AI IDE, Cursor, 
\textbf{(2)} a quantitative and qualitative analysis of the design issues detected by static analysis tools (i.e., SonarQube and CodeScene) in the projects generated by Cursor, \textbf{(3)} a curated dataset of 10 project descriptions for large-scale project generation, 169,646 lines of Cursor-generated code across 10 projects, and the design issues identified in these projects, and \textbf{(4)} recommendations and suggestions for practitioners on adopting AI IDEs effectively to avoid design issues.

\textbf{Paper Organization}: Section~\ref{sec:relatedwork} discusses related work pertaining to this study. Section~\ref{sec:rqdataset} presents our research questions and outlines the methodology employed for our data collection (including project generation and design issue detection). We provide our study results in Section~\ref{sec:results}, followed by a discussion of the findings and implications in  Section~\ref{sec:discussion}. We then clarify the potential threats to the validity of this study in Section~\ref{sec:validity} followed by the conclusions in Section~\ref{sec:conclusion}. 

\section{Related Work}\label{sec:relatedwork}
Several studies have examined various aspects of AI coding tools, including LLM-based coding assistants, AI IDEs and coding agents, design quality evaluation of the generated code, and end-to-end project generation.

\textbf{LLM-based Code Generation Tools.}
A large body of work has investigated LLM-based code generation tools, including AI coding assistants such as GitHub Copilot and general-purpose LLMs such as ChatGPT, studying various aspects of AI-generated code, including quality, correctness, and productivity. Li \textit{et al.} analyzed 2,547 shared ChatGPT conversations from GitHub using the DevChat dataset \cite{li2025unveiling}. They identified Code Generation \& Completion as the most prominent task supported by ChatGPT. Ouyang \textit{et al.} studied the non-determinism of ChatGPT in code generation by evaluating the solutions to 829 code generation problems and measuring the semantic, syntactic, and structural similarities of the generated code. They reported that ChatGPT exhibits a high degree of non-determinism, highlighting challenges for reproducibility and reliability in LLM-based code generation \cite{ouyang2024empirical}. Similarly, Jin \textit{et al.} conducted an empirical study using the DevGPT dataset, analyzing real-world conversations between developers and ChatGPT to understand how developers use it for code generation. They found that ChatGPT is primarily used to demonstrate high-level concepts, and that the generated code is often not directly production-ready, highlighting its limited usefulness for real-world software development \cite{jin2024can}. Liang \textit{et al.} conducted a large-scale survey of 410 developers to understand usability challenges in AI code generation tools. They found that developers primarily use these tools to reduce keystrokes, speed up programming, and recall syntax to improve their productivity, but often struggle to control the tools to generate the desired output \cite{UsabilityOfAIProgramming}. Similarly, Barke \textit{et al.} conducted a user study with 20 programmers interacting with GitHub Copilot across multiple programming tasks. They reported that developers' interaction with Copilot is bimodal: \textit{acceleration mode} when they know the solution and use the tool to speed up coding, and an \textit{exploration mode} when they are uncertain and use the tool to explore ideas for a solution \cite{barke2023grounded}. Dakhel \textit{et al.} studied GitHub Copilot by assessing its ability to solve programming problems and comparing its generated solutions with human-written code. Their results show that, although Copilot can generate solutions for almost all programming problems, these solutions may contain bugs. They concluded that Copilot can be an asset for experienced developers but a liability for novice developers who struggle to identify suboptimal code \cite{dakhel2023github}. Fu \textit{et al.} analyzed Copilot-generated code from real-world GitHub projects using static analysis tools to detect security weaknesses \cite{fu2025security}. The study's results revealed a high likelihood of security vulnerabilities in Copilot-generated code (e.g., 29.5\% in Python and 24.2\% in JavaScript). 

\textbf{AI IDEs and Agent-based Coding Tools.} A new generation of coding tools, including AI IDEs and coding agents (e.g., Cursor~\cite{Cursor}, Claude Code~\cite{Claude}, and OpenAI Codex~\cite{Codex}), has recently been introduced and has drawn considerable attention from both researchers and practitioners. Li \textit{et al.} mined GitHub pull requests from five coding agents and introduced the AIDEV dataset~\cite{li2025rise}. Their results show the rapid adoption of coding agents in open-source projects. They also found that, while coding agents significantly accelerate code generation and developer productivity, their pull requests are less frequently accepted, and the generated code tends to be structurally simpler. Similarly, He \textit{et al.} conducted an empirical study to estimate the causal effect of adopting Cursor on development velocity and software quality~\cite{he2025speed}. They compared Cursor-adopting GitHub projects with similar projects that do not use Cursor and found that Cursor adoption leads to a significant but transient increase in development velocity, while causing a persistent rise in code complexity and static analysis warnings. Becker \textit{et al.} conducted a randomized controlled trial with experienced open-source developers to evaluate the impact of early-2025 AI coding tools (e.g., Cursor Pro) on developers' productivity~\cite{becker2025measuring}. They compared developer productivity with and without AI assistance using task-based experiments and found that AI assistance does not improve productivity for experienced developers on complex tasks and can even slow them down. Huang \textit{et al.} conducted an empirical study on AI-generated and human-written pull requests by analyzing code quality, maintainability, and reviewer reactions~\cite{huang2026more}. Their results show that pull requests generated by LLM agents tend to exhibit significantly higher code redundancy compared to human-written ones, and reviewers often express more neutral or positive sentiments toward AI-generated contributions.

\textbf{Static Analysis for Design Quality Evaluation.} Several studies have used static code analysis tools to evaluate AI-generated code for design issues and code smells. For instance, Liu \textit{et al.} evaluated 4,066 ChatGPT-generated programs across Java and Python for correctness and quality issues~\cite{liu2024refining}. They found that while a substantial portion of the generated code is correct, many programs contain incorrect outputs, compilation and runtime errors, and maintainability issues detected via static analysis. They further demonstrated that iterative refinement using ChatGPT can partially mitigate these issues, improving code quality by over 20\%. Similarly, Sabra \textit{et al.} used SonarQube, a static analysis tool, to evaluate the quality and security of AI-generated code generated by multiple LLMs using 4,442 Java programming tasks~\cite{sabra2025assessing}. They reported that although LLMs often generate functionally correct code, it frequently contains bugs, security vulnerabilities, and code smells, including critical issues such as hard-coded credentials and path traversal vulnerabilities. Santa Molison \textit{et al.} compared LLM-generated and human-written code by analyzing Python solutions across multiple difficulty levels and evaluating them with SonarQube \cite{santa2025llm}. Their results show that LLM-generated code generally contains fewer bugs and requires less effort to fix, with fine-tuned models reducing high-severity issues. However, in more complex tasks, LLM-generated solutions may introduce structural problems not observed in human-written code.

\textbf{End-to-End Project Generation.} Several studies have explored end-to-end project generation. Hong \textit{et al.} proposed MetaGPT, a novel framework that incorporates efficient human workflows into LLM-based multi-agent collaborations. Their results show that MetaGPT achieved higher \textit{Pass@1} scores than prior multi-agent approaches, exceeding 85\% on HumanEval and MBPP benchmarks~\cite{hong2023metagpt}. They also evaluated MetaGPT through human evaluation on the SoftwareDev dataset, including challenging end-to-end project-generation tasks, thereby demonstrating MetaGPT's autonomous code generation capabilities. Nguyen \textit{et al.} presented AgileCoder, a multi-agent system that integrates Agile Methodology by assigning roles (e.g., Product Manager, Developer, Tester) to develop software from user inputs using sprints for incremental delivery~\cite{nguyen2025agilecoder}. They evaluated AgileCoder against prior multi-agent development frameworks and reported higher \textit{Pass@1} scores than ChatDev and MetaGPT on HumanEval and MBPP benchmarks. They also evaluated AgileCoder on the ProjectDev dataset, which contains end-to-end project generation tasks, and reported improved results compared to ChatDev and MetaGPT. Wan \textit{et al.} proposed TDDev, a multi-agent framework based on Test-Driven Development (TDD) to automatically generate full-stack Web applications from natural language requirements~\cite{wan2025automatically}. Their approach employs multiple LLM-based agents to decompose requirements, generate executable test cases, produce frontend and backend code, simulate user interactions, and iteratively refine the implementation until requirements are fulfilled. They reported that TDDev achieved improvement in overall accuracy in generating full-stack applications by 14.4\% compared to state-of-the-art baselines.

\textbf{Comparative Summary.} (1) Existing literature has not explored AI IDEs such as Cursor for large-scale project generation. Most prior work on end-to-end project generation proposes frameworks for generating software projects; however, these projects are relatively small and are generated from simple project descriptions (e.g., ``\textit{Create a Python program to develop an interactive weather dashboard}'') \cite{hong2023metagpt, nguyen2025agilecoder, wan2025automatically}. In contrast, our study proposes the FD-HITL framework for systematically generating large-scale projects using detailed and self-curated project descriptions. (2) Prior studies have extensively evaluated snippet-level code generated by LLM-based coding assistants using static analysis to assess code quality \cite{sabra2025assessing, santa2025llm, liu2024refining}, but have not examined the design quality of project-level code generated using AI IDEs. (3) Existing research on AI IDEs and agent-based coding tools mostly consists of empirical studies analyzing PR-level agent contributions from open-source GitHub repositories \cite{li2025rise, he2025speed, huang2026more}. In contrast, our study focuses on the design quality of end-to-end large-scale projects generated by an AI IDE (i.e., Cursor).

\section{Research Methodology}\label{sec:rqdataset}
\begin{figure}[h]
    \centering
    \includegraphics[width=1\linewidth]{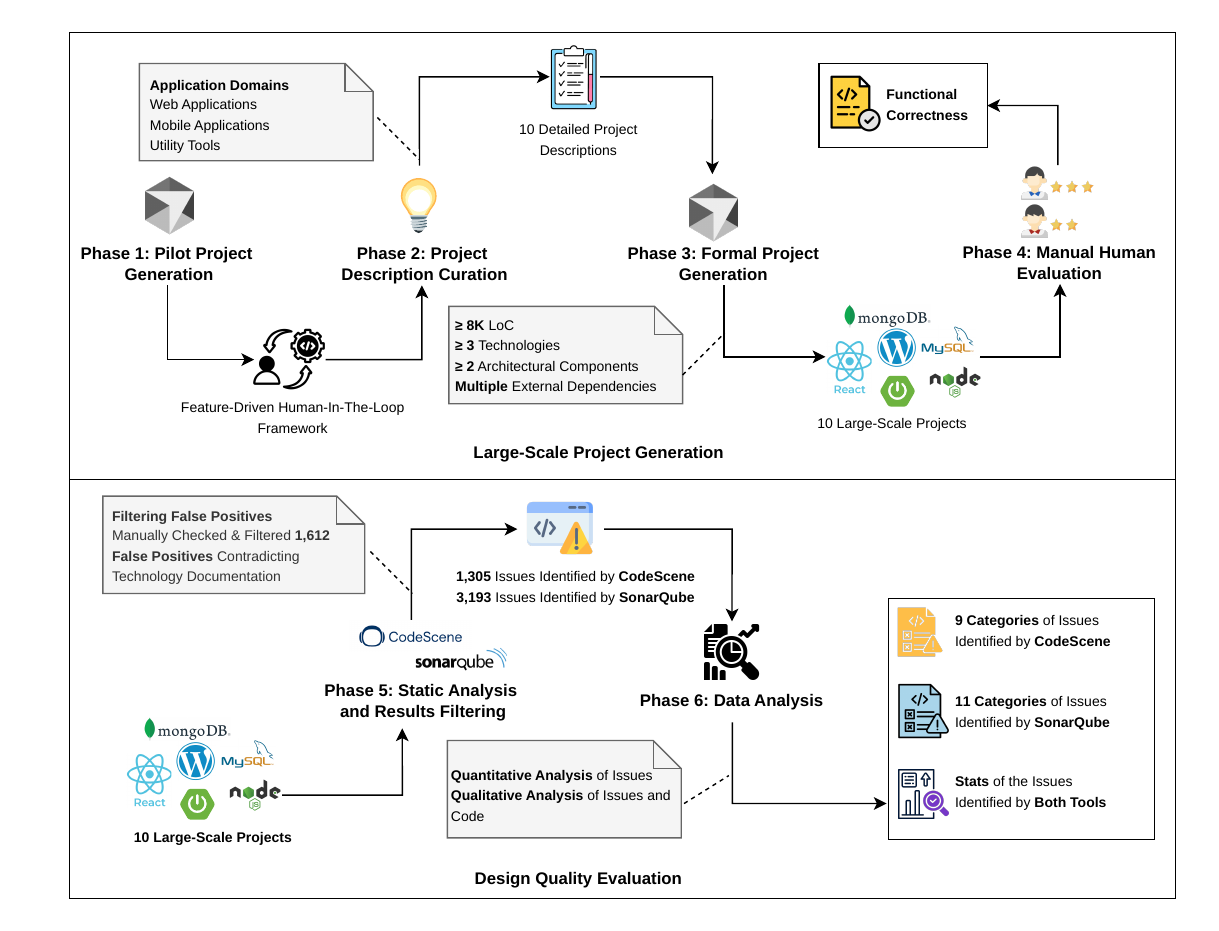}
    \caption{\textcolor{black}{Overview of our research process}}
    \label{fig:methodology}
    \vspace{-1em}
\end{figure}

Our empirical study comprises two phases: (1) a generation of 10 complex large-scale projects using Cursor from 10 systematically curated project descriptions, and (2) a design evaluation of the generated projects to identify common design issues in these projects. Figure~\ref{fig:methodology} shows an overview of our research methodology.

\subsection{Research Goal and Questions}\label{sec:goals}

Our study \textit{aims} to \textit{explore Cursor to generate large-scale projects} and \textit{evaluate the design quality of the Cursor-generated projects}. We first \textit{propose a framework for generating large-scale projects}, and then \textit{evaluate the generated projects for design issues and understand the nature of the identified issues}. 

We formulated two main RQs to achieve this goal. RQ1 evaluates the proposed Feature-Driven Human-In-The-Loop (FD-HITL) framework, which comprises strategies to effectively generate large-scale projects using Cursor. RQ1 also explores the extent to which Cursor enables developers to build large-scale projects. Finally, RQ2 identifies and analyzes the design issues in Cursor-generated projects. Particularly, we address the following Research Questions (RQs):

\textbf{RQ1}: \textit{To what extent can Cursor generate large-scale projects?}

\textbf{Motivation.} AI-powered IDEs (e.g., Cursor) have demonstrated the ability to generate project-level code within the project context. This RQ explores large-scale project generation using Cursor to demonstrate the extent to which it can generate such projects.

\textbf{RQ1.1:} \textit{What are the characteristics of the large-scale projects generated by Cursor?}

\textbf{Motivation}. This RQ aims to study the characteristics of large-scale projects generated using Cursor, including project size, application domain, architectural style, architectural components, technology stack, and the number of design issues. These characteristics are important for determining whether Cursor can generate projects of similar scale and complexity to industrial projects.

\textbf{RQ1.2}: \textit{How effective is the FD-HITL framework in generating functional large-scale projects using Cursor?}

\textbf{Motivation}. This RQ aims to evaluate the effectiveness of the Feature-Driven Human-In-The-Loop (FD-HITL) framework in generating functionally correct large-scale projects. Currently, developers rely on ad hoc prompting strategies that may not scale to complex projects. Therefore, a systematic framework is needed to effectively generate large-scale projects.

\textbf{RQ2}: \textit{What design issues are found in large-scale projects generated by Cursor?}

\textbf{Motivation}. Cursor can autonomously generate project-level code across multiple files within the project context, unlike previous AI code generation tools, which could only generate code snippets \cite{li2025rise, lu2026projdevbench}. This autonomous generation of a large amount of code can introduce high-level (method and file-level) design issues in projects that may differ from those found in AI-generated code snippets. This RQ focuses on evaluating the design quality of large-scale Cursor-generated projects. We define separate sub-RQs for CodeScene and SonarQube because these two tools identify design issues at different levels of granularity.    

\textbf{RQ2.1}: \textit{What design issues are identified by CodeScene in large-scale projects generated by Cursor?}

\textbf{Motivation}. Based on our pilot study results, CodeScene and SonarQube tools identify different issues. CodeScene mainly identifies issues at a high level of code, such as functional complexity and code duplication. This RQ focuses on the results of CodeScene to identify a wide range of design issues. 

\textbf{RQ2.2}: \textit{What design issues are identified by SonarQube in large-scale projects generated by Cursor?}

\textbf{Motivation}. Our pilot study results show that SonarQube also identifies issues at a low level, such as variable assignment and type issues. This RQ focuses only on the results of SonarQube to identify a diverse set of design issues.

\textbf{RQ2.3} \textit{What are the key overlapping design issues identified by both CodeScene and SonarQube?} 

\textbf{Motivation}. CodeScene and SonarQube detect issues differently. However, overlapping issues between these tools suggest more critical design issues. This RQ aims to identify the overlapping issues to highlight critical design issues confirmed by both tools.  

\textbf{RQ2.4}: \textit{What technology-specific design issues does Cursor consistently generate in large-scale projects?}

\textbf{Motivation}. Different technology stacks have unique conventions and best practices (e.g., React state management, Spring Boot dependency injection). This RQ identifies which issues are specific to certain technologies, helping developers and tool builders anticipate and address technology-specific issues when using Cursor.



\subsection{Pilot Project Generation} \label{subsec:Pilotstudy}
We began our pilot study by generating projects using the SoftwareDev dataset provided by MetaGPT~\cite{hong2023metagpt}. 
However, we found that the project descriptions generated by MetaGPT were relatively simple, resulting in small projects with few LoC (less than 1k). While SonarQube identified some issues, such as naming convention violations and dead code, these findings were not particularly substantial. CodeScene, on the other hand, failed to detect any significant issues. Based on these initial results, we concluded that more extensive project descriptions are necessary to produce larger projects. Consequently, we decided to curate our own project descriptions to facilitate the generation of large-scale projects inspired by the common project ideas in Web and mobile application development \cite{WebIdeas, MobileIdeas}.

We curated eight project descriptions for the subsequent phase of our pilot study. We provided these descriptions to Cursor and interacted with Cursor based on its code output. However, we were only able to generate basic features with a limited size. As projects grew in size (in terms of LoC and files), they either failed to run or contained significant logical errors. After generating pilot projects, we executed them for logical issues and any possible errors. We also evaluated these projects manually and found that the generated projects did not fully meet the requirements or expectations listed in the project descriptions. 
At this stage, we identified a limited number of issues using our analysis tools.

The pilot study yielded two main insights. First, detailed project descriptions are essential for generating large projects to detect meaningful design issues. Second, we identified key factors influencing Cursor's performance in generating large-scale projects, including structured planning, decomposition of projects into low-level tasks, separation of frontend, backend, and database tasks, feature-wise incremental and iterative development, human-in-the-loop feedback, and continuous testing.

\subsection{Feature-Driven Human-In-The-Loop (FD-HITL) Framework} \label{subsec:framework}
Based on the pilot study, we found that the ad hoc prompting approach (e.g., \textit{Vibe Coding} \cite{fawzy2025vibe}) is not appropriate for large-scale project generation. Our pilot study shows some key limitations: (1) The ad hoc prompting approach addresses a complex problem in one go and lacks the decomposition of the high-level project description into independently testable features and low-level tasks, (2) It lacks incremental and iterative development with constant human feedback, and (3) It misses the validation of the independently testable features. Based on these observations, we proposed a systematic framework for generating large-scale projects using Cursor, shown in Figure \ref{fig:framework}. We adapted the Feature-Driven Development (FDD) process \cite{palmer2001practical} for generating our projects. This framework is divided into four distinct phases, explained as follows:

\begin{figure}[h]
    \centering
    \includegraphics[width=0.8\linewidth]{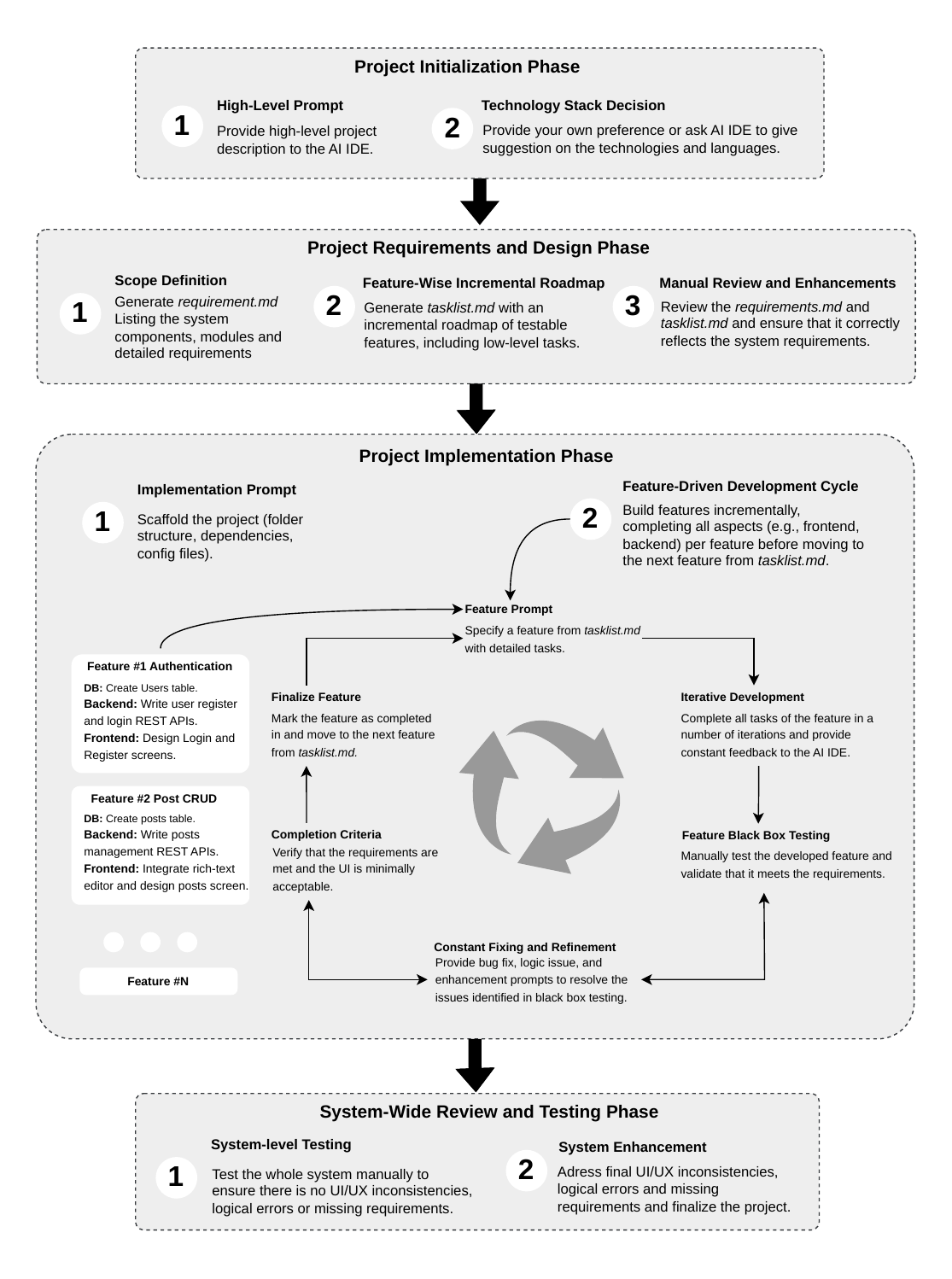}
    \caption{Overview of the Feature-Driven Human-In-The-Loop (FD-HITL) Framework}
    \label{fig:framework}
    \vspace{-1em}
\end{figure}

\noindent\textbf{Project Initialization Phase.} The initialization phase establishes the project's foundation by explaining the project context and selecting an appropriate technology stack through collaboration with Cursor. This phase consists of two sequential steps: (1) a \textit{high-level prompt} and (2) a \textit{technology stack decision}.

In the first step, we provide Cursor with a \textit{high-level prompt} structured around two key components. First, we establish the project context by specifying the business domain (e.g., Chinese vocabulary learning) and provide structured requirements to Cursor through an informal specification of system features and user workflows. Second, we provide process instructions that guide Cursor’s approach, such as directing it to plan before generating code to avoid premature implementation. In the second step, we engage in collaborative reasoning with Cursor, focused on technology stack selection.  We either ask Cursor to suggest a technology stack that best fits the project description, or we provide our own choice based on our past experience and allow Cursor to reason about and compare it with alternatives to facilitate an informed decision. An example of a high-level prompt is shown in Prompt \ref{box:initialization}, which includes the project description, process instructions, and a suggested technology stack.

\begin{promptbox}[box:initialization]{An example prompt of the initialization phase}
\textbf{[Project Description: \href{https://github.com/Kashifraz/DIinAGP/blob/main/projects/Project_Descriptions.pdf}{see examples here}]}\\

This is the high-level project description of the project we want to build. We will build this project iteratively and in multiple increments. Please do not directly jump to writing code. We will first do planning and then start writing code. \\

Can we use Java Spring Boot with (Maven) for backend services and React Native for Frontend mobile app?
\end{promptbox}

\noindent\textbf{Project Requirements and Design Phase.} This phase comprises three main steps. First, we define the project scope by prompting Cursor to generate detailed system requirements from the informal software specification (e.g., the project description) provided in the \textit{high-level prompt}. In this step, Cursor generates the \texttt{requirements.md} file specifying the system components, modules, and detailed requirements. Secondly, we prompt Cursor to create a list of independently testable features by grouping all related database, backend, and frontend tasks. Cursor generates a \texttt{tasklist.md} file detailing a \textit{feature-wise incremental roadmap} for the project development (see Prompt \ref{box:design}). This roadmap decomposes the entire project into testable features, enabling Cursor to generate code incrementally and iteratively. In the last step, we manually review the \texttt{requirements.md} and \texttt{tasklist.md} files and make corrections where needed.

\begin{promptbox}[box:design]{An example prompt of Project Requirements and Design Phase}

Create two files in the root directory 1) requirements.md 2) tasklist.md \\

The requirements.md will serve as the single source of truth for the system architecture and functional requirements of the Vocabulary App. It should define all major components, their modules, and the detailed requirements for each module. \\

The tasklist.md should provide the feature-wise Incremental development roadmap, listing features in the order they should be built. Each section should represent a complete, independent feature, with all related database, backend, and frontend tasks grouped together. We will only focus on the database, backend, and frontend tasks.
\end{promptbox}

\noindent\textbf{Project Implementation Phase.} In this phase, we first prompt Cursor to generate the project structures of all system components (e.g., backend Spring Boot, frontend React Native) and ensure that these project structures are running without any errors. We provide an \textit{implementation prompt} (see Prompt \ref{box:implementation}) to Cursor to scaffold project structures and configure the project environment, including configuration files and dependencies. In this step, we also focus on establishing the connection to the database, creating a testing backend API, and a test screen to help verify that project structures are working without errors. 

\begin{figure}[htbp]
\begin{promptbox}[box:implementation]{Example prompt of the Implementation Phase}
\textbf{Backend project structure setup.} You can now start working on Project Setup. Please first focus on the backend project setup inside the backend directory. Please do not jump to the functionality, but only create the working project setup with a test API endpoint, and also establish the database connection.\\

\textbf{Frontend project structure setup.} The backend project is now working fine. Please start working on the frontend project setup now. Please do not jump to UI screens and functionality, just create a running frontend project with a test screen.
\end{promptbox}
\end{figure}

After generating the project structure, we enter the \textit{feature-driven development cycle} (see Prompt \ref{box:cycle}). (1) In this cycle, we pick a feature from the \texttt{tasklist.md} file and iteratively complete all tasks for that feature with constant human feedback. (2) For every feature, we first complete the backend and database tasks and test the generated resources (e.g., backend APIs). Once the backend APIs are working, we move to the frontend tasks to integrate them and design the frontend UI screens. To test the backend APIs, we use either cURL commands or the Postman API testing tool. (3) Once all backend and frontend tasks are completed, we manually tests the feature's functionality. (4) We provide \textit{bug-fix}, \textit{logical-issue}, or \textit{enhancement prompts} to resolve the issues identified during testing of the feature. For effective debugging, we use Cursor features such as attaching terminal error messages to the prompt and taking screenshots, which help Cursor identify the cause of the issue. We also use other debugging strategies, including providing browser console logs and network tab responses in the prompt to Cursor. (5) We verify that the requirements for this feature are fulfilled, and there are no remaining issues. (6) Finally, we mark the feature as completed and move to the next feature in the \texttt{tasklist.md} file. This \textit{feature-driven development cycle} continues until all the features listed in \texttt{tasklist.md} file are completed.  

\begin{promptbox}[box:cycle]{Example prompts of Feature-driven Development Cycle}
\textbf{Database and backend tasks of a feature.} Feature 1 is complete. Now, please start working on Feature 2: User Authentication (detailed in tasklist.md). Please first focus on the database and backend tasks. Then, we will test the backend API endpoints before moving to the frontend tasks.\\

\textbf{Frontend tasks of a feature.} The backend APIs are working. Now, please work on integrating these backend APIs on the frontend side and complete the frontend tasks. Please ensure that the UI design is according to modern UI principles.
\end{promptbox}

\noindent\textbf{System-Wide Review and Testing Phase.} Once the \textit{feature-driven development cycle} is complete, we inform Cursor that all features have been implemented and that we will now test and review the entire system. This phase consists of two steps. In the \textit{system-level testing} step, we manually test the entire system against the specifications in the \texttt{requirements.md} file to ensure that all requirements are met and that the UI does not have any inconsistencies. Finally, in \textit{system enhancement} step, we make any necessary improvements, such as implementing missing requirements, fixing logical errors, or improving the UI. We provide \textit{logical-issue} or \textit{enhancement prompts} to resolve identified issues and ensure that the system works as expected (see Prompt \ref{box:review}). Finally, we mark the project as complete.

\begin{promptbox}[box:review]{Example prompts of Review and Testing Phase}
\textbf{Marking project as completed and enhancement prompt.} All features are now complete. Now focus on improving the whole application. First, we need to improve the design of the bottom tab navigation. The current design is very basic. We need modern bottom navigation. \\

\textbf{Logical-issue prompt.} The preview on the form edit and create pages is not accurate and needs to be fixed. On the actual page, the form appears differently, but the preview on the edit or create pages does not show it properly. I have attached screenshots of the actual form and the preview of the edit or create pages.
\end{promptbox}

\subsection{Project Description Curation}
Based on the pilot study, we curated 10 complex project descriptions for our large-scale projects. We targeted projects in the following application domains: Web applications, utility tools, and mobile applications. We chose these application domains because they are popular among developers who use AI-generated code~\cite{kashif2025developers}. These project descriptions are designed to include complex tasks, such as database operations (e.g., CRUD), complex business logic, user interface (UI) requirements, and the use of external libraries or dependencies. The first author curated the initial project descriptions, and other two co-authors reviewed them and provided feedback. These project descriptions are inspired by the common project ideas in Web and mobile application development \cite{WebIdeas, MobileIdeas}. 

\subsection{Formal Project Generation} \label{subsec:formalgeneration}
After finalizing the project descriptions, we generated the formal 10 large-scale projects using Cursor Pro with ``automatic LLM model selection'' to balance generation speed and code quality. We generated 9 projects in a client-server architecture style and only one in a monolithic layered architecture style, a WordPress plugin (P6\_FormPlugin). The choice of programming languages depends on the selected technology stacks for our Web, mobile, and tool application domains. These projects were built using popular technologies (see Table \ref{tab:projectstats}) associated with JavaScript, Java, Python, and PHP languages. The first author generated the projects without writing any code manually, but followed the FD-HITL framework to systematically guide Cursor and provided continuous feedback. The first author only manually set up the development environment for these projects to run without errors.

We included projects in our dataset only when they met our inclusion and exclusion criteria, as detailed in Table \ref{tab:projectcriteria}. Since the projects built during the pilot study were small and less complex, we did not include them in our dataset for the analysis of design issues. To demonstrate our criteria, we present the included \texttt{P2\_VocabularyApp} project as an example: (1) The project comprises 11,159 LoC across 84 source files. We calculated physical LoC using the \texttt{cloc}\footnote{\url{https://github.com/AlDanial/cloc}} command-line tool. We excluded third-party dependencies (e.g., \texttt{node\_modules/}, \texttt{target/}), configuration files (e.g., \texttt{.yaml}, \texttt{.properties}), and documentation files (e.g., \texttt{.md}) to count only the code directly generated by Cursor. (2) It uses at least three technologies: Java Spring Boot (backend), React Native (mobile frontend), and MySQL (database). (3) It follows client-server architecture with RESTful API communication. (4) It uses several external dependencies, including \texttt{Flyway} for database schema migrations and \texttt{React Navigation} for managing navigation in a React Native frontend application. This demonstrates that the project meets the inclusion criteria (e.g., I1, I2, I3, and I4) and is valid for inclusion in our dataset. These criteria ensured that our dataset included large-scale projects with complexity comparable to that of industrial software systems \cite{SO-survey2025, soto2021comprehensive, latendresse2022not}. The source code for the 10 Cursor-generated projects is available in our replication package in the \texttt{projects} folder \cite{replicationPackage}.

\begin{table}[htbp]
\footnotesize
\setlength{\tabcolsep}{3pt} 
\caption{Inclusion and exclusion criteria for the AI-IDE generated projects}
\label{tab:projectcriteria}
\renewcommand{\arraystretch}{0.9} 
\begin{tabular}{>{\raggedright\arraybackslash}p{0.05\linewidth}>{\raggedright\arraybackslash}p{0.85\linewidth}}
\toprule
\multicolumn{2}{l}{\textbf{Inclusion Criteria}} \\
\midrule
I1 & The project should have at least 8k lines of code (LoC). \\
I2 & The project should be generated in a technology stack with at least 3 technologies. \\
I3 & The project is built in a complex architectural style with multiple architectural components. \\ 
E4 & The project uses external dependencies. \\
\midrule
\multicolumn{2}{l}{\textbf{Exclusion Criteria}} \\
\midrule
E1 & The project is small and not built in a complex architectural style with multiple architectural components. \\ 
E1 & The project does not use external dependencies. \\
\bottomrule
\end{tabular}
\end{table}

\subsection{Evaluation Process} \label{evaluation_process}
We conducted a manual human evaluation to assess the functionality of projects generated with Cursor. Manual evaluation was performed by executing the generated large-scale projects and then manually reviewing them against the system requirements in \texttt{requirements.md} file. The main goal was to determine whether the generated projects executed successfully and met the requirements defined in the \texttt{requirements.md} file. Similar human evaluation methods have been used in the existing literature to assess code generated by AI coding tools. For instance, Hong \textit{et al.} used human evaluation to assess the projects generated by their proposed MetaGPT framework (including demo projects such as games, tools, and small applications), using metrics such as Executability and Human Revision Cost \cite{hong2023metagpt}. 
Similarly, Nguyen \textit{et al.} employed human evaluation to assess Executability and Requirements Satisfaction of software generated by their AgileCoder multi-agent framework \cite{nguyen2025agilecoder}.

We executed the generated project to verify its functionality against the \texttt{requirements.md} file. We manually tested the projects against each requirement listed in the \texttt{requirements.md} file and then labeled the requirement either as ``\textit{Complete}'' or ``\textit{Incomplete}''. Two authors completed this manual process for all large-scale projects and calculated their functional correctness. We made functional correctness verifiable by capturing screenshots of all \textit{complete} functional requirements, available in our replication package \cite{replicationPackage}. This analysis shows the extent to which Cursor can generate functional large-scale projects.

\subsection{Static Analysis and Results Filtering} \label{subsec:staticanalysis}
We used two well-known automated static analysis tools, CodeScene \cite{CodeScene} and SonarQube \cite{SonarQube}, widely used in industry and research for identifying software quality issues~\cite{borg2024ghost,tornhill2022code,santa2025llm, etemadi2022sorald,yu2023towards}. Prior work characterizes code and design smells as poor solutions to recurring implementation and design problems that degrade software quality and hinder system evolution by making changes more difficult to implement \cite{moha2009decor, Design2014, yamashita2012code}. Therefore, we treat both code and design smells identified by CodeScene and SonarQube as design issues in this study.

\subsubsection{CodeScene Analysis.} We employed CodeScene's static analysis features via its IDE extension to identify design issues in projects generated using Cursor. We selected the CodeScene IDE extension for two reasons: (1) it detects and highlights code issues directly within the IDE, streamlining the analysis and export of design issues, and (2) it provides comprehensive design issue detection necessary for addressing RQ2. We did not use CodeScene Community Edition, which offers advanced analysis features useful only for projects with historical version control data. Our large-scale projects lack version-control histories because we generated them using Cursor for evaluation purposes. We installed the CodeScene IDE extension within our Cursor environment (Cursor Pro), where it operated continuously in the background, detecting design issues throughout project generation. Upon completion of each project, we retrieved the list of design issues from the problems tab and exported it to a spreadsheet for analysis. We completed the CodeScene analysis of our 10 projects and identified a total of 1,305 design issues.

\subsubsection{SonarQube Analysis.} \label{subsubsec:sonarmethod} We used SonarQube Free Server edition and configured the analysis with the default rules to detect design issues in our projects. We used the SonarQube web interface to create projects and ran the Sonar Scanner to detect code issues. 
Since our projects contain multiple architectural components (e.g., frontend and backend), we ran the Sonar Scanner separately from the root directory of each architectural component. The same web interface was used to visualize the detected design issues. We used SonarQube REST API to export the issues to spreadsheet files and included these files in our replication package \cite{replicationPackage}.

The tool identified 4,805 design issues in the 10 generated projects, which we then manually verified by examining the issue descriptions and confirming them in the code. Our manual verification identified a large number of recurring false positives. We classified an issue as a false positive only when it clearly contradicted the documentation of the project's underlying technology. For example, SonarQube flagged the issue ``\textit{Rename class} \texttt{Form\_Plugin\_Admin} \textit{to match the regular expression} \texttt{\^{}[A-Z][a-zA-Z0-9]*\$}'', a naming convention issue that discourages underscores in \texttt{class} names. However, this issue appeared in \texttt{P6\_FormPlugin}, a WordPress Plugin Project, where using underscores in \texttt{class} names is a recommended practice to avoid naming collisions according to WordPress documentation\footnote{\url{https://developer.wordpress.org/plugins/plugin-basics/best-practices/}}. As a result of our filtration process, we identified 1,612 (33.5\%) recurring false-positive issues across 8 distinct issue patterns and removed them from our dataset. An issue pattern is a consolidated representation of recurring design issues. 

After filtering out false positives, we ended up with 3,193 valid design issues detected by SonarQube. A significant number of these issues recur across the 10 projects and occur in two different ways: (1) issues with identical descriptions repeated across projects, and (2) issues with similar underlying problems but varying specific instances. A detailed manual analysis of these 3,193 issues is both time-consuming and inefficient. Therefore, we merged these recurring issues into unique issue patterns. For example, SonarQube identified ``\textit{Unexpected negated condition}'' 51 times across the 10 projects, which we consolidated into a single issue pattern with the same issue name. Similarly, SonarQube identified cognitive complexity issues 82 times across the projects, with descriptions such as ``\textit{Refactor this method to reduce its Cognitive Complexity from 23 to the 15 allowed}'' and ``\textit{Refactor this function to reduce its Cognitive Complexity from 34 to the 15 allowed}''. We merged these into a single issue pattern: ``\textit{Refactor this method to reduce its Cognitive Complexity from ... to the 15 allowed}''. Through this process, we reduced the 3,193 issues to 142 unique issue patterns, eliminating duplication and facilitating qualitative analysis.  

\subsection{Data Analysis} \label{subsec:dataanalysis}
We employed quantitative and qualitative analysis techniques to analyze the issues detected by the scanning tools, as explained below.

\subsubsection{Quantitative Analysis} 

We identified 1,305 design issues using CodeScene and 3,193 design issues using SonarQube across the 10 projects. We created separate spreadsheet files for both tools to list the design issues identified by them. For each project, we created an individual spreadsheet file containing two sheets: ``\textit{Issues}'' and ``\textit{Stats}''. The ``\textit{Issues}'' sheet lists all design issues detected by the tools, while the ``\textit{Stats}'' sheet summarizes the unique issues and their occurrence frequency within the project. Note that the ``\textit{unique issues}'' in the case of SonarQube refer to consolidated ``\textit{issue patterns}'', as the issues identified by SonarQube are highly recurring and therefore grouped into unique issue patterns (see Section \ref{subsubsec:sonarmethod}). We also created a ``\texttt{Summary.xlsx}'' file that consolidates all unique issues across all 10 projects. This summary file shows: (1) the frequency of each unique issue per project, (2) the total frequency of each unique issue across all 10 projects, (3) the total number of issues identified in each project, and (4) the overall total of identified issues across all 10 projects (e.g., 1,305 CodeScene and 3,193 SonarQube issues). Table \ref{tab:projectissues} presents the number of design issues detected by both tools.

\begin{table}[htbp]
\centering
\small
\caption{Number of issues found by SonarQube and CodeScene across different projects}
\label{tab:projectissues}
\begin{tabular}{@{}p{0.1\linewidth}p{0.2\linewidth}p{0.2\linewidth}p{0.2\linewidth}@{}}
\toprule
\textbf{Project ID} & \textbf{Project Name} & \textbf{SonarQube Issues} & \textbf{CodeScene Issues} \\ \midrule
P1 & CVBuilder & 231 & 217\\
P2 & VocabularyApp & 350 & 90\\
P3 & Ecommerce & 268 & 113\\
P4 & JobApplication & 253& 75 \\
P5 & ChartMaker & 53 & 62\\
P6 & FormPlugin & 97& 74 \\
P7 & BlogWebsite & 584 & 116 \\
P8 & SocialApp & 572 & 168\\
P9 & LMS & 383 & 199\\
P10 & POS & 402& 191\\ \bottomrule
\end{tabular}
\end{table}

We manually checked the overlap between issues identified by both tools only at a similar level of granularity to answer RQ2.3. For instance, we observed overlap between the issue pattern identified by SonarQube, ``\textit{Refactor this function to reduce its Cognitive Complexity from ... to the 15 allowed},'' and the \textit{Complex Method} design issues identified by CodeScene, as both are identified at the method level. To address RQ2.4, we further categorized the identified issues based on whether they are technology-specific or general (i.e., applicable across all technologies). All issues identified by CodeScene were general. However, we identified 38 issue patterns (1,900 issues) from SonarQube that were technology-specific. For example, the issue pattern ``\textit{`...' is missing in props validation}'' is specific to React. We also grouped issues related to similar technologies together. For instance, we grouped React, Vanilla JavaScript, and Node.js issues under the ``\textit{JavaScript/React/Node.js}'' category, as they belong to the JavaScript ecosystem. The complete list of these issue patterns from SonarQube is provided in the \texttt{Technology\_Specific\_Issues.xlsx} file. These spreadsheet files are available in our replication package, organized into separate folders for issues identified by CodeScene and SonarQube \cite{replicationPackage}. 

\subsubsection{Qualitative Analysis} We qualitatively analyzed only the design issues identified by SonarQube to group them into high-level categories. In contrast, for CodeScene, we directly used the categories provided by the tool, as they are already defined at a high level. We employed thematic analysis as our primary methodology to analyze the identified issues. We adopted a predominantly inductive and data-driven approach to provide a comprehensive interpretation of the data~\cite{braun2006using}. This method allowed us to systematically identify, analyze, and organize patterns in the design issues detected by the static analysis tools.

We performed the initial coding by identifying and labeling the core concepts of the issue pattern from the issue descriptions and the corresponding source code in the projects. We manually checked two to three example issues for each issue pattern to code all 142 issue patterns due to the recurring nature of issues identified by SonarQube. The first author carefully read each issue description and the corresponding code snippets to extract code. We iteratively refined the codes as we examined more issue patterns. At the same time, the other two co-authors engaged in iterative review and discussion to enhance the reliability of the coding process \cite{McDonald2019reliability}. This approach aligns with established qualitative research practices, in which a single primary coder conducts initial coding.

We developed high-level categories by iteratively comparing the codes identified during the initial coding phase. We developed 11 high-level categories from the qualitative analysis of issues identified by SonarQube. For instance, we categorized the issue, ``\textit{Refactor this code to not nest functions more than 4 levels deep}'' into the \textit{Cognitive Complexity} category. SonarQube flagged this issue on \texttt{addNodeView()} function in \href{https://github.com/Kashifraz/DIinAGP/blob/main/projects/P7_BlogWebsite/admin_frontend/src/extensions/AuthorBoxExtension.jsx#L132}{\texttt{AuthorBoxExtension.jsx}} (\texttt{P7\_BlogWebsite}). This function contains nested callback functions for more than 4 levels deep, which increases the cognitive complexity of the code. Similarly, we categorized the issue ``\textit{Constructor has 20 parameters, which is greater than 7 authorized}'' into the \textit{Design Principle Violation} category. SonarQube flagged this issue in \href{https://github.com/Kashifraz/DIinAGP/blob/main/projects/P3_Ecommerce/backend/src/main/java/com/ecommerce/dto/OrderResponse.java#L44}{\texttt{OrderResponse.java}} (\texttt{P3\_Ecommerce}). One reason for categorizing this issue as a design principle violation is that a constructor with many parameters suggests that the class likely has multiple responsibilities or is tightly coupled. The first author conducted the initial coding, while the other two co-authors reviewed it and provided feedback on these categories. The calculated pairwise Cohen’s kappa between the primary coder and the reviewing co-authors was 0.88, indicating substantial agreement. We resolved disagreements using the negotiated agreement approach \cite{campbell2013coding} to enhance the reliability of our data analysis. 

After completing the qualitative analysis of design issues identified by SonarQube and obtaining the design issue categories from both SonarQube and CodeScene (the latter provided by the tool), we mapped these categories to widely recognized design principles. For instance, we mapped the \textit{Complex Method} and \textit{Large Method} design issues as violations of the \textit{Single Responsibility Principle (SRP)}. We demonstrate this with an example method, \href{https://github.com/Kashifraz/DIinAGP/blob/main/projects/P5_ChartMaker/frontend/src/components/ChartPreview.vue#L343}{\texttt{renderChart()}} (\texttt{P5\_ChartMaker}), which is detected as both \textit{Complex Method} and \textit{Large Method}. This method handles multiple responsibilities, including validation, chart initialization and rendering, DOM style manipulation, and resize handling. Similarly, we mapped \textit{Complex Conditionals} (e.g., in \href{https://github.com/Kashifraz/DIinAGP/blob/main/projects/P6_FormPlugin/admin/js/submissions.js#L172}{\texttt{submissions.js}} (\texttt{P6\_FormPlugin})) identified by CodeScene and \textit{Control Flow and Conditionals Issues} (e.g., in \href{https://github.com/Kashifraz/DIinAGP/blob/main/projects/P10_POS/backend/app/Controllers/Api/ExpenseController.php#L147}{\texttt{ExpenseController.php}} (\texttt{P10\_POS})) identified by SonarQube as violations of the \textit{Keep It Simple, Stupid (KISS)} principle, which emphasizes that unnecessary complexity should be avoided.

\section{Results}\label{sec:results}

\subsection{RQ1: To what extent can Cursor generate large-scale projects?} \label{subsec:resultRQ1}

\subsubsection{RQ1.1: What are the characteristics of the large-scale projects generated by Cursor?} \label{subsubsec:resultRQ1.2}

In this section, we present the characteristics (e.g., LoC, files, technology stack) of the large-scale projects generated by Cursor. These characteristics illustrate the extent to which AI IDEs can enable developers to build large-scale projects by following a systematic approach.

\textbf{Project size}. We report the size of our large-scale projects in terms of LoC and files per project, as shown in Table \ref{tab:projectstats}. The minimum LoC is 8,594 in the \texttt{P6\_FormPlugin} project, while the maximum is 28,505 LoC in \texttt{P10\_POS}, with an average of 16,965 LoC. In terms of files, the smallest count is 22 files in \texttt{P6\_FormPlugin}, while the largest is 233 files in \texttt{P9\_LMS}. On average, each project contains 114 files.

\textbf{Application domains and technology stacks}. The application domains of the 10 projects include 2 mobile applications, 4 Web applications, and 4 utility tools. We built these projects using technology stacks widely adopted in industry for these application domains \cite{SO-survey2025}, as presented in Table \ref{tab:projectstats}. For example, for the frontends of both mobile applications (\texttt{P2\_VocabularyApp} and \texttt{P8\_SocialApp}), we used React Native, while the backend services were implemented with Spring Boot in Java and a MySQL database. Similarly, we developed three Web-based systems using JavaScript's MERN stack (\underline{M}ongoDB, \underline{E}xpress, \underline{R}eact, and \underline{N}ode.js). In addition, we also used Vue.js and React for frontend development, alongside Java Spring Boot, Python Django, and PHP CodeIgniter for building RESTful APIs for the backend. 
One project, the \texttt{P6\_FormPlugin}, was developed using the WordPress Plugin API with PHP. 

\textbf{Architectural components}. Eight projects in our dataset have three main architectural components: \textit{frontend}, \textit{backend}, and \textit{database}. \texttt{P7\_BlogWebsite} has four architectural components, including an \textit{admin\_frontend}, a \textit{user\_frontend}, a \textit{backend}, and a \textit{database}. \texttt{P6\_FormPlugin} has a monolithic WordPress \textit{Plugin} component and a \textit{database} component.

\textbf{External dependencies}. We also recorded the external dependencies, including third-party libraries and development dependencies, listed in the configuration files (e.g., \texttt{package.json}, \texttt{pom.xml}) used by the projects (see Table \ref{tab:projectstats}). The maximum number of dependencies is 47, observed in \texttt{P1\_CVBuilder}, with a mean of 30 dependencies per project. This reflects similarities to industrial projects, in which external dependencies are commonly used to implement complex functionality \cite{soto2021comprehensive,latendresse2022not}. For example, \texttt{P8\_SocialApp} uses the \texttt{rn-emoji-keyboard} library to provide an emoji selection interface and \texttt{react-navigation} library to implement drawer and bottom tab navigation in the mobile application. Similarly, \texttt{P7\_BlogWebsite} uses the \texttt{Tiptap} library to build a rich-text editor and the \texttt{axios} library for HTTP communication with backend RESTful API endpoints.

\begin{table}[htbp]
\centering
\small
\caption{Overview of large-scale projects generated by Cursor}
\label{tab:projectstats}
\begin{tabular}{p{0.16\linewidth}p{0.07\linewidth}p{0.05\linewidth}p{0.13\linewidth}p{0.13\linewidth}p{0.32\linewidth}}
\toprule
\textbf{Project} & \textbf{LoC} & \textbf{Files} & \textbf{Functional Correctness}& \textbf{Dependencies} & \textbf{Technology Stack} \\ \midrule
 P1\_CVBuilder& 26,798 & 137 & 86.36\%& 47 & MERN stack \\
 P2\_VocabularyApp& 11,066 & 83 & 84.61\%& 35 & React Native, Spring Boot \& MySQL \\
 P3\_Ecommerce& 14,644 & 112 & 88.46\%& 37 & React, Spring Boot \& MySQL \\
 P4\_JobApplication& 15,312 & 79 & 89.47\%& 28 & MERN stack \\
 P5\_ChartMaker& 9,391 & 49 & 92.85\%& 20 & Vue.js, Django \& MySQL \\
 P6\_FormPlugin& 8,594 & 27 & 92.85\%& 4 & PHP, MySQL \& WordPress API \\
 P7\_BlogWebsite& 15,071 & 77 & 95.23\%& 44 & MERN stack with Next.js \\
 P8\_SocialApp& 12,545 & 116 & 96.15\%& 44 & React Native, Spring Boot \& MySQL \\
 P9\_LMS& 27,720 & 233 & 93.33\%& 24 & Vue.js, Spring Boot \& MySQL \\
 P10\_POS& 28,505 & 226 & 93.93\%& 19 & Vue.js, CodeIgniter \& MySQL \\ \bottomrule
\end{tabular}

\end{table}

\begin{tcolorbox}[
    colback=lightgray!20, 
    colframe=darkgray,   
    boxrule=0.5mm,        
    arc=2mm,              
    title=Finding 1,      
    fonttitle=\bfseries   
]
When used with the systematic FD-HITL framework, Cursor generated 10 large-scale projects, with an average of 16,965 LoC and 114 files per project, built on widely used technology stacks and external dependencies.
\end{tcolorbox}

\subsubsection{RQ1.2: How effective is the FD-HITL framework in generating functional large-scale projects using Cursor?}
We used human evaluation to assess the functional correctness of the projects generated by Cursor against the \texttt{requirements.md} files. We manually executed the project and verified the system against each requirement listed in the \texttt{requirements.md} file (see Section \ref{evaluation_process}). We also captured screenshots of the \textit{complete} functional requirements, which have been made available in the replication package~\cite{replicationPackage}. The value of functional correctness of the 10 projects is also reported in Table~\ref{tab:projectstats}. The results of the human evaluation indicate that Cursor can assist developers in generating large-scale projects when following a systematic FD-HITL framework.

\begin{figure}[h]
    \centering
    \includegraphics[width=1\linewidth]{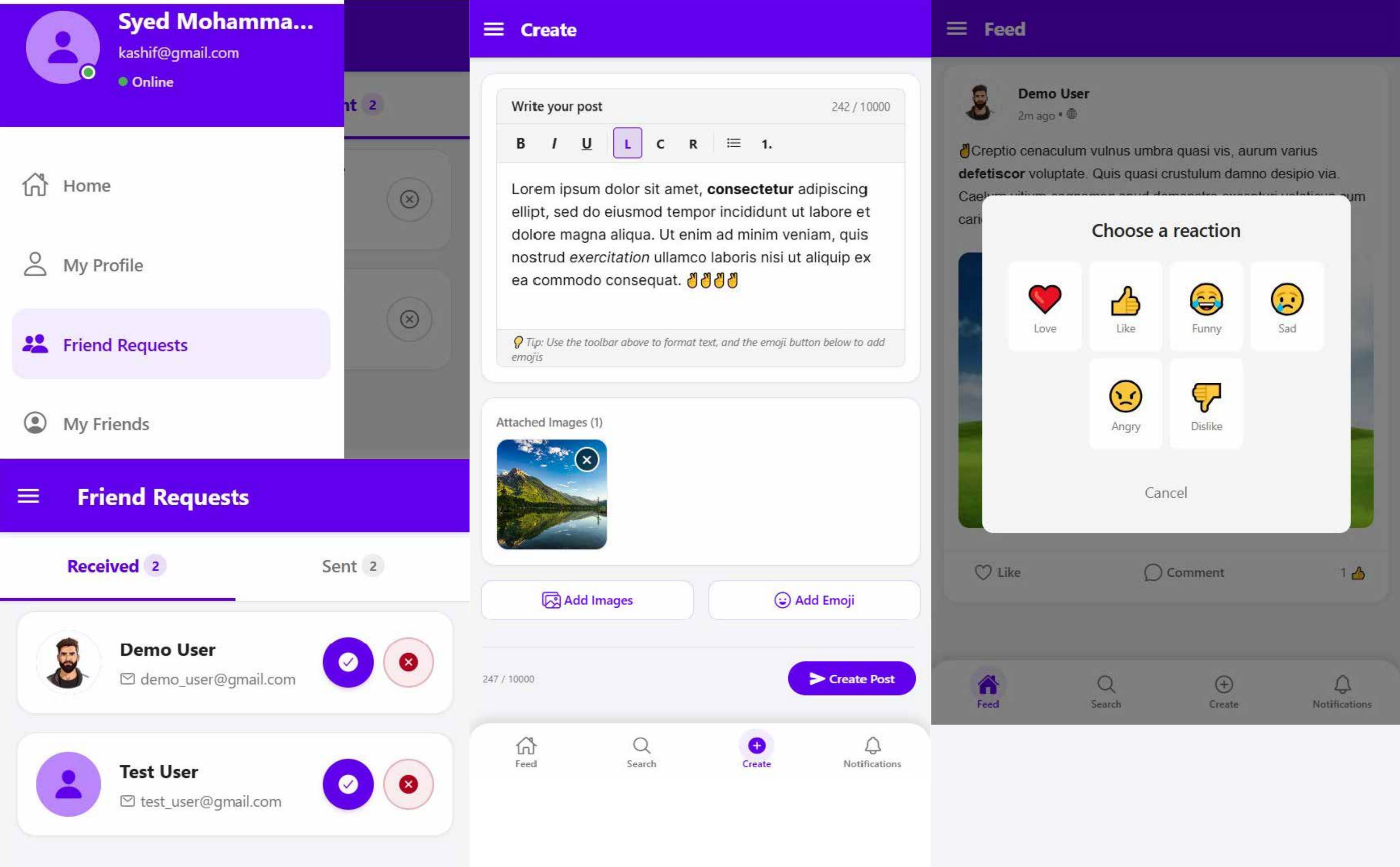}
    \caption{An example of a mobile application (\texttt{P8\_SocialApp}) generated using Cursor}
    \label{fig:socialapp}
    \vspace{-1em}
\end{figure}

To demonstrate the functional correctness of Cursor-generated projects, we present examples from two projects: \texttt{P8\_SocialApp} and \texttt{P7\_BlogWebsite}. The \texttt{P8\_SocialApp} project includes 26 functional requirements as stated in the \texttt{requirements.md} file, which was initially generated by Cursor and subsequently reviewed and modified by the first author. We demonstrate a subset of representative screenshots of the \texttt{P8\_SocialApp} project in Figure~\ref{fig:socialapp}. Our manual evaluation of the \texttt{P8\_SocialApp} project shows 96.2\% functional correctness, with only one functional requirement marked as \textit{Incomplete} during the evaluation process. We marked the requirement as \textit{Incomplete} when the functionality is either missing completely or not properly implemented due to logical errors. As shown in Figure~\ref{fig:socialapp}, we successfully built a running mobile application by following the FD-HITL framework. For instance, Cursor successfully implements UI features such as drawer navigation and bottom tab navigation. Similarly, core functionalities, including image upload, emoji selection, post creation, reaction and comment systems, and friend management, are correctly implemented and functional. Only one functional requirement (post-update) is not fully met based on the manual evaluation: users cannot update their posts once created. 

\begin{figure}[h]
    \centering
    \includegraphics[width=1\linewidth]{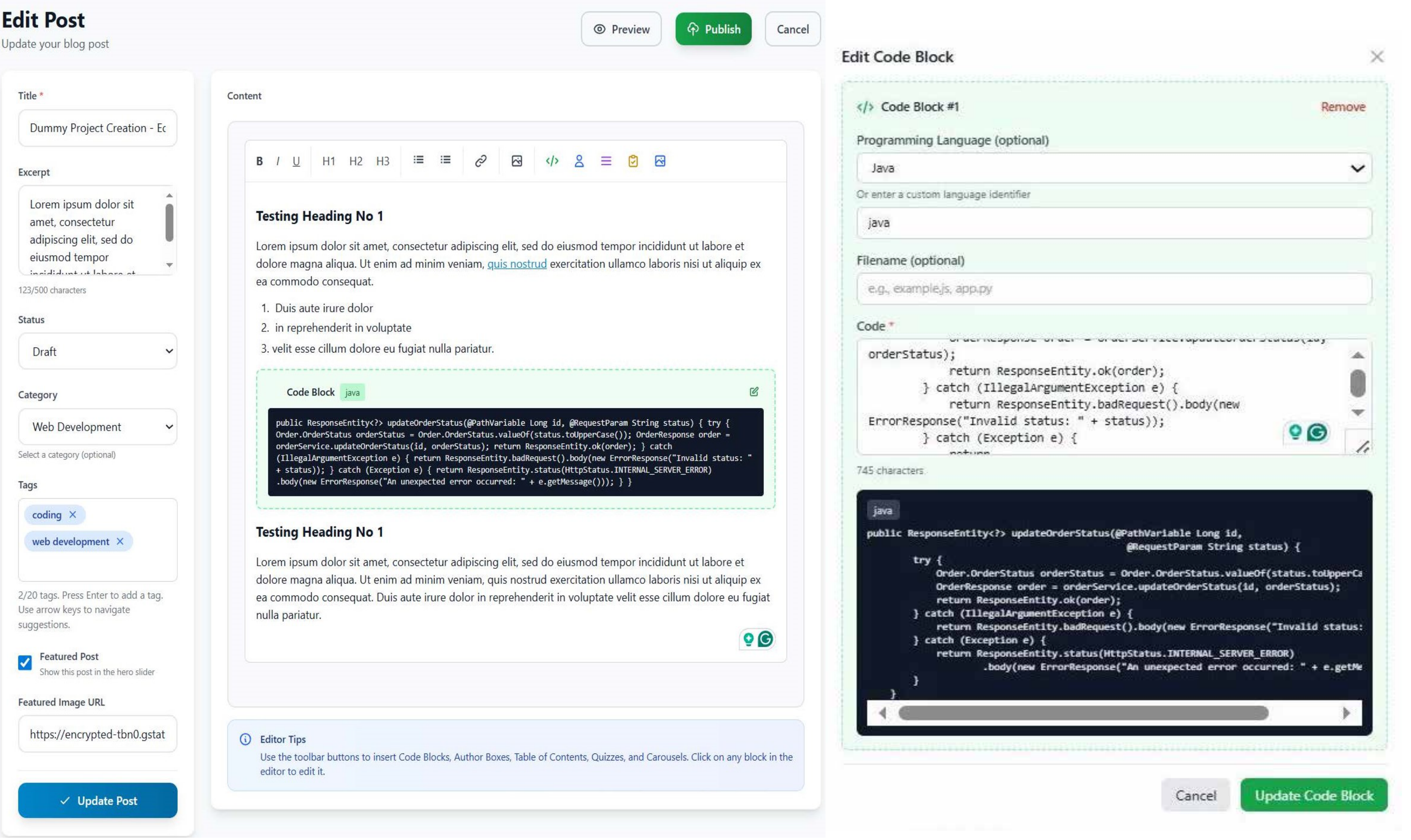}
    \caption{An example of a Web application (\texttt{P7\_BlogWebsite}) generated using Cursor}
    \label{fig:blogwebsite}
    \vspace{-1em}
\end{figure}

The \texttt{P7\_BlogWebsite} project has 21 main functional requirements. Our manual evaluation of this project shows 95\% functional correctness. We provide a few example screenshots illustrating the functionality of the \texttt{P7\_BlogWebsite} project in Figure~\ref{fig:blogwebsite}. As shown in the figure, Cursor successfully builds a running Web application and implements functionality such as creating post content, providing post metadata, adding content using customizable blocks (e.g., code snippet blocks or author boxes), updating post content, and publishing posts for public visibility. Overall, 20 out of 21 functional requirements are evaluated as \textit{complete}.

The manual evaluation of the 10 projects against their intended functional requirements yields an average functional correctness of 91\%, with the highest at 96\% and the lowest at 85\%. The highest functional correctness is observed in \texttt{P8\_SocialApp}, with 96.2\% of requirements marked as \textit{complete}. These results suggest that the FD-HITL framework can help developers build large-scale projects using AI IDEs. However, our manual evaluation also identified 16 \textit{incomplete} requirements, including 11 missed requirements and 5 logical issues that caused incomplete functionality. For instance, \texttt{P8\_SocialApp} has one \textit{incomplete} requirement due to a missing requirement: the absence of post-update functionality. Similarly, \texttt{P7\_BlogWebsite} has an incomplete requirement caused by a logical issue in the rich-text editor, which does not properly render bold and italic fonts.

\begin{tcolorbox}[
    colback=lightgray!20, 
    colframe=darkgray,   
    boxrule=0.5mm,        
    arc=2mm,              
    title=Finding 2,      
    fonttitle=\bfseries   
]
Cursor can generate functionally correct large-scale projects when used with a systematic framework, with an average functional correctness of 91\% across the 10 projects. However, human evaluation also found some missing or incompletely implemented requirements in the generated projects.  
\end{tcolorbox}
 
\subsection{RQ2: What design issues are likely to appear in large-scale projects generated by Cursor?}

\subsubsection{RQ2.1: What design issues are identified by CodeScene in large-scale projects generated by Cursor?} \label{subsubsec:codescene}
\begin{figure}[h]
    \centering
    \includegraphics[width=1 \linewidth]{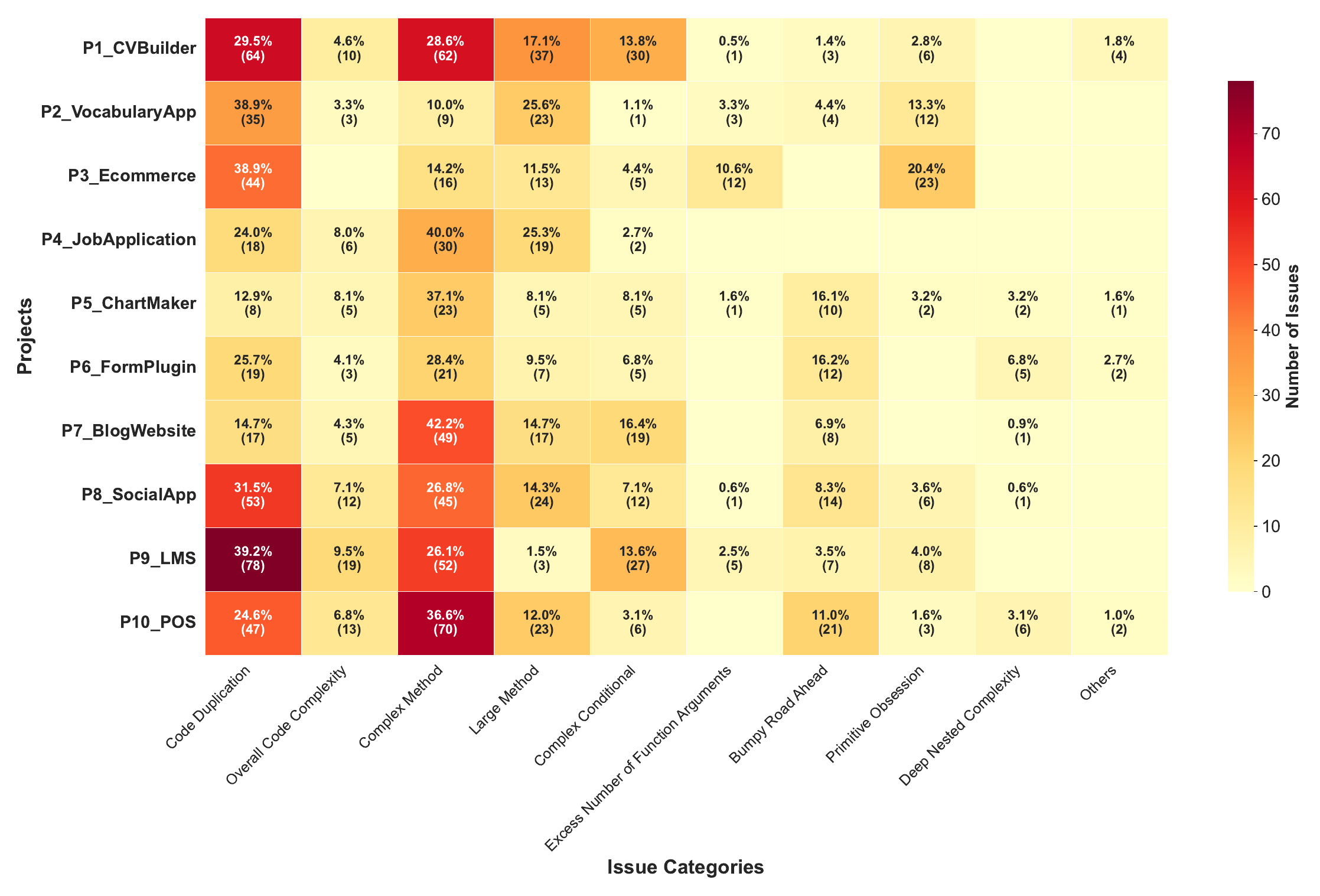}
    \caption{Overview of the categories of the design issues identified by CodeScene}
    \label{fig:codescene}
    \vspace{-1em}
\end{figure}

We employed CodeScene to identify design issues in the 10 large-scale projects generated by Cursor. We identified 1,305 design issues using CodeScene, categorizing them into 9 categories (see Figure \ref{fig:codescene}). The mapping of design issues identified by CodeScene to design principles is shown in Table \ref{tab:principle-mapping}.

\begin{table}[htbp]
\centering
\footnotesize
\caption{Mapping of design principles to issues identified by CodeScene and SonarQube }
\label{tab:principle-mapping}
\begin{tabular}{p{0.2\linewidth}p{0.35\linewidth}p{0.35\linewidth}}
\toprule
\textbf{Principle} & \textbf{CodeScene}& \textbf{SonarQube}\\
\midrule
\textit{Don’t Repeat Yourself (DRY)} \cite{fowler2008repetition} & Code Duplication & Dead, Duplicate or Redundant Code \\
\addlinespace
\textit{Single Responsibility Principle (SRP)} \cite{martin2014srp} & Complex Method,  Large Method, Excess Number of Function Arguments& Design Principle Violation \\
\addlinespace
\textit{Separation of Concerns (SoC)} \cite{tarr1999ndegrees} & Overall Code Complexity& \\
\addlinespace
\textit{Fail Fast} \cite{shore2004fail} & & Exception-Handling Issue  \\
\addlinespace
\textit{Keep It Simple, Stupid (KISS)} \cite{martin2009clean} & Complex Conditional, Bumpy Road Ahead, Deep Nested Complexity & Code Complexity, Control Flow and Conditionals Issue\\
 \bottomrule
\end{tabular}
\end{table}


\textbf{(1)} \textit{Code Duplication} \textbf{383 (28.4\%).} This category of issues indicates duplicated or copy-pasted code within the source code files. Such duplication makes the code harder to maintain (e.g., it increases the cost of change, as the same logical change must be applied in multiple places). Code duplication design issues are most frequently detected by the CodeScene tool and violate the \textit{Don’t Repeat Yourself (DRY)} \cite{fowler2008repetition} design principle. Quantitative analysis of \textit{Code Duplication} design issues shows that they occur slightly more frequently in frontend code (208, 54.3\%) than in backend code (175, 45.7\%). We present an example from the \href{https://github.com/Kashifraz/DIinAGP/blob/main/projects/P2_VocabularyApp/backend/src/main/java/com/vocabularyapp/service/QuizGenerationService.java#L66}{\texttt{QuizGenerationService.java}} (\texttt{P2\_VocabularyApp}). This file is responsible for generating quizzes with three different difficulty levels: easy, medium, and hard. This single file contains six methods that largely duplicate the same logic, making the code repetitive. Specifically, the methods \texttt{generateEasyModeQuiz()}, \texttt{generateMediumModeQuiz()}, and \texttt{generateHardModeQuiz()} repeat nearly identical logic with only minor variations. Similarly, the methods \texttt{generateMultipleChoiceOptions()}, \texttt{generateMediumModeOptions()}, and \texttt{generateHardModeOptions()} also exhibit substantial code duplication. These six methods contain approximately 150 duplicated LoC and only 30 unique LoC, which pose several maintenance challenges, such as: (1) a single bug fix requires changes in three different locations, (2) feature additions require separate code in three different methods, and (3) the code requires three times more test cases.

\begin{figure}[h]
    \centering
    \includegraphics[width=1 \linewidth]{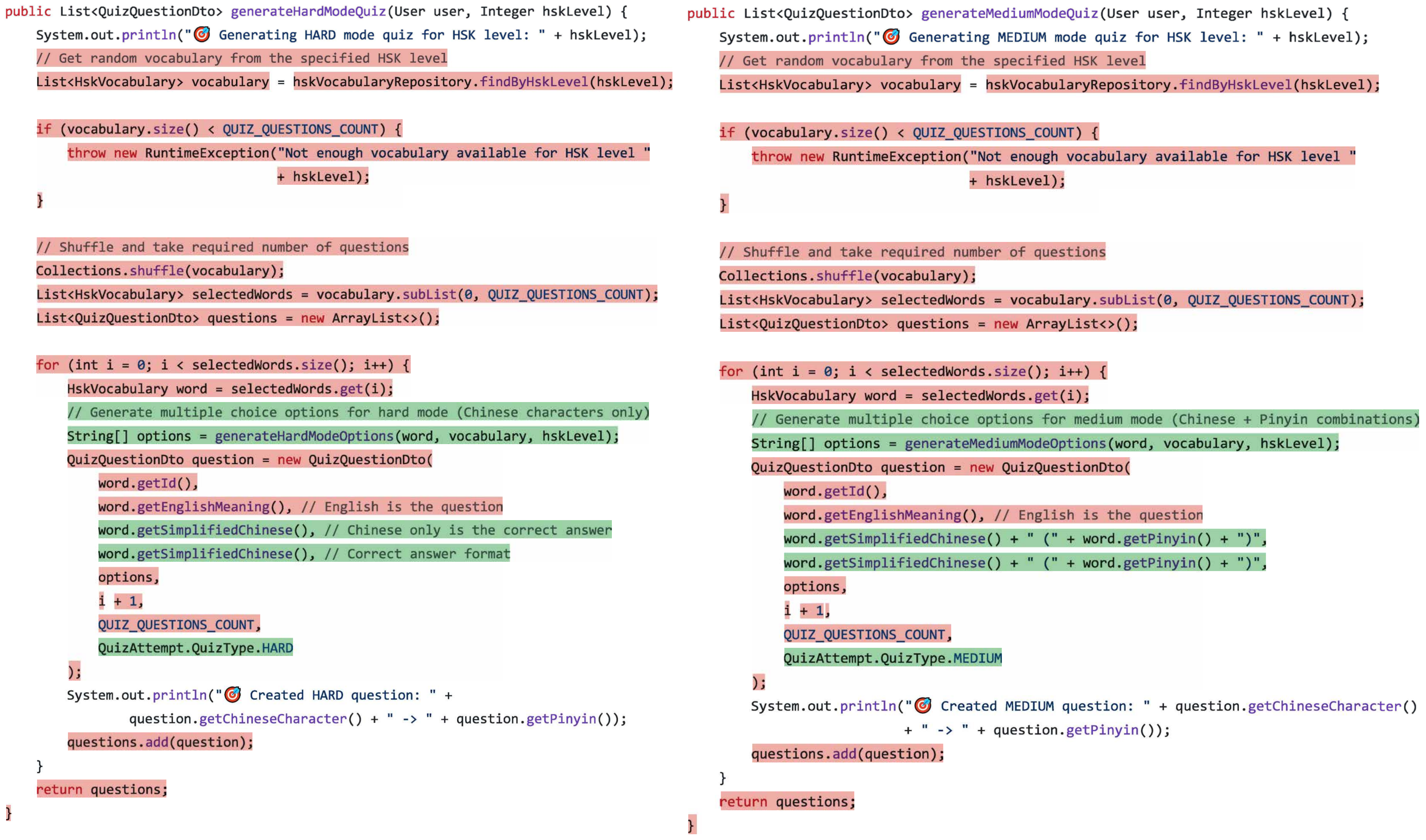}
    \caption{An example of \textit{Code Duplication} design issue}
    \label{fig:duplication}
    \vspace{-1em}
\end{figure}

\begin{figure}[h]
    \centering
    \includegraphics[width=0.4\textwidth]{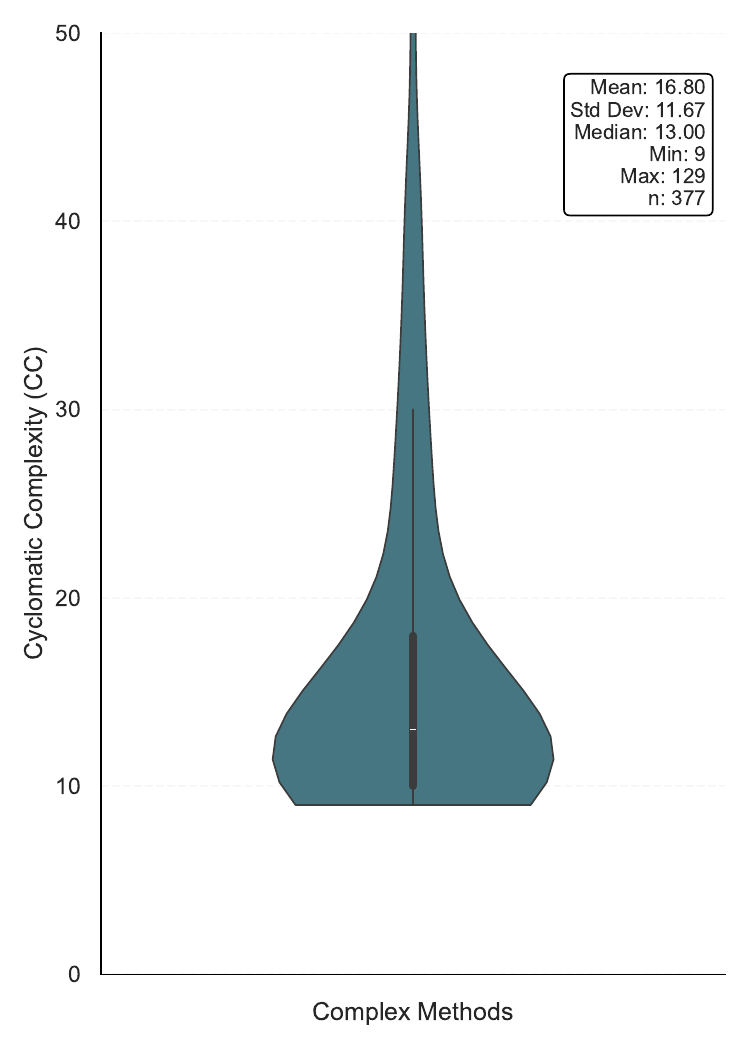}
    \hfill
    \includegraphics[width=0.4\textwidth]{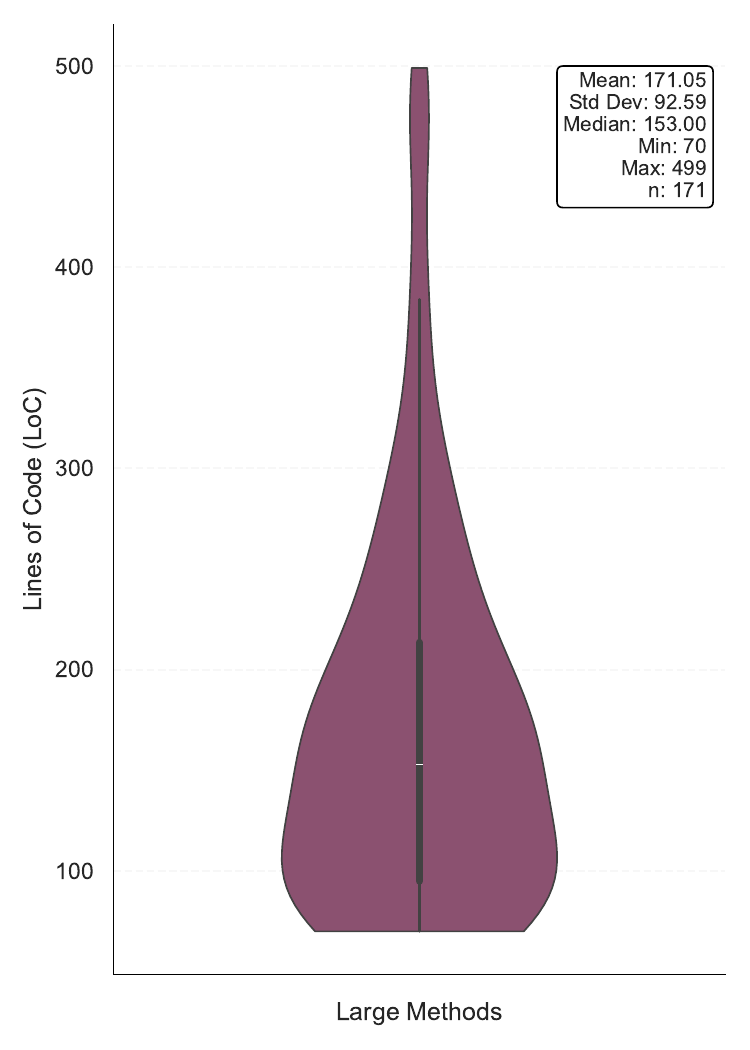}    
    \caption{Distribution of cyclomatic complexity (CC) in \textit{Complex Method} design issues and lines of code (LoC) in \textit{Large Method} design issues in Cursor-generated projects}
    \label{fig:largecomplex}
\end{figure}

\textbf{(2)} \textit{Complex Method} \textbf{377 (27.9\%)}. This category of design issues includes methods with high \textit{cyclomatic complexity}, which measures code complexity by counting the number of linearly independent logical paths through the code, thereby indicating the difficulty of the code in terms of understanding, testing, and maintenance. After quantitatively analyzing \textit{Complex Method} design issues, we found that they occur more frequently in frontend code (247, 66.5\%) than in backend code (130, 35.5\%). We illustrate \textit{Complex Method} design issues using \href{https://github.com/Kashifraz/DIinAGP/blob/main/projects/P5_ChartMaker/backend/api/views.py#L280}{\texttt{views.py}}  (\texttt{P5\_ChartMaker}). This file contains a \texttt{post()} method within the \texttt{DataIngestionView} class, which has a high cyclomatic complexity of 36. This method implements a RESTful API endpoint that parses user-uploaded data files to create charts and stores a corresponding data table record in the database. The high cyclomatic complexity arises because the method includes nine \texttt{if-else} statements, five \texttt{try-except} blocks, and multiple exit points, with nine \texttt{return} statements. The primary cause of this complexity is that the method handles multiple responsibilities, including file retrieval, data processing, and creating data table records, thereby violating the \textit{Single Responsibility Principle (SRP)} \cite{martin2014srp}. As a result, the method is difficult to test, as the large number of logical paths requires a substantial number of test cases to achieve adequate code coverage. The \textit{cyclomatic complexity} distribution of the 377 \textit{Complex Method} design issues identified across our 10 projects is illustrated in Figure~\ref{fig:largecomplex}. Most \textit{Complex Methods} have cyclomatic complexity values in the range of 10 to 20, with a mean value of approximately 17. However, a considerable number of methods exhibit cyclomatic complexity values above 20.

\textbf{(3)} \textit{Large Method} \textbf{171 (12.6\%)}. These design issues arise when a method contains excessive code, making it difficult to understand, maintain, and test. \textit{Large Methods} can also hide logical errors and increase the likelihood of code duplication. To demonstrate the \textit{Large Method} design issues, we present the example of the \texttt{getApplicable()} method in \href{https://github.com/Kashifraz/DIinAGP/blob/main/projects/P10_POS/backend/app/Controllers/Api/DiscountController.php#L306}{\texttt{DiscountController.php}} (\texttt{P10\_POS}). This method uses a RESTful API endpoint to retrieve applicable discounts based on the transaction amount. The method contains 206 LoC, which is large for a single method. The high LoC arises because the method performs multiple responsibilities, including user authentication and validation, transaction handling, discount querying, and result processing. This violates the \textit{SRP}, which states that a method should ideally perform a single, well-defined task. In addition, the method includes 19 log statements for debugging, which further increase its length. Figure~\ref{fig:largecomplex} shows the LoC distribution of the 171 \textit{Large Methods} identified across our 10 projects. Most of these \textit{Large Methods} fall within the 100 to 200 LoC range, with a mean value of 171. Interestingly, 73.7\% (123) of the \textit{Large Method} design issues are found in the frontend implementations, while 26.3\% (48) are found in the backend code.  

\begin{figure}[h]
    \centering
    \includegraphics[width=0.8 \linewidth]{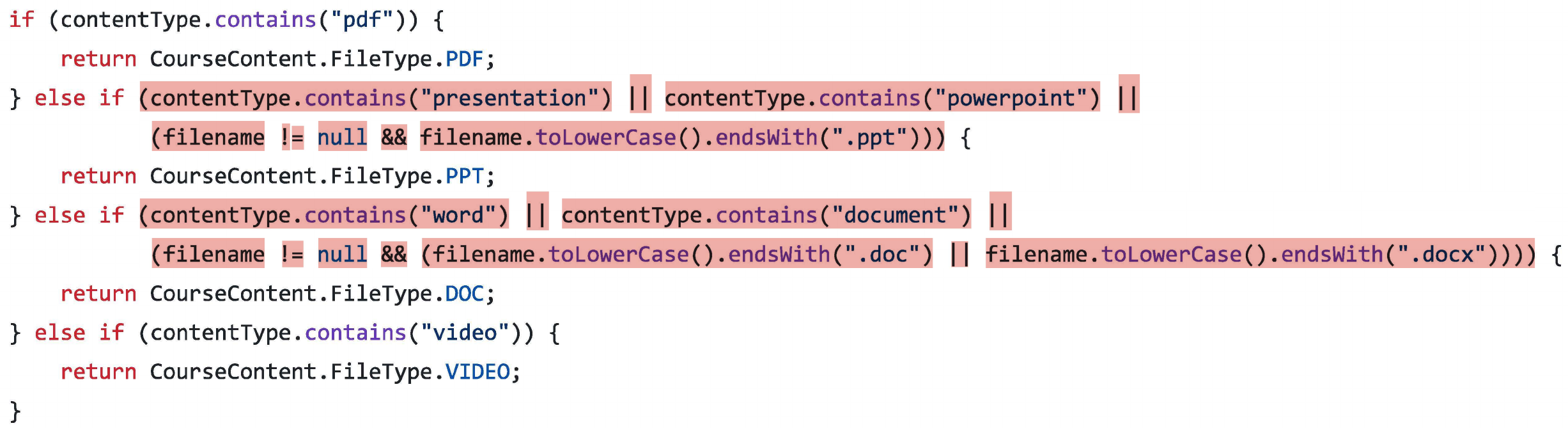}
    \caption{Examples of \textit{Complex Conditional} design issue}
    \label{fig:complexconditional}
    \vspace{-1em}
\end{figure}

\textbf{(4)} \textit{Complex Conditional} \textbf{112 (8.2\%)}. These design issues indicate that a conditional expression is overly complex and difficult to understand. A \textit{Complex Conditional} refers to an expression within a branch, such as an \texttt{if} statement, that consists of multiple logical operations, unlike \textit{Complex Method} design issues, which are identified based on \textit{cyclomatic complexity}. A quantitative analysis of \textit{Complex Conditional} design issues shows that 79 (70.5\%) of these design issues occur in the frontend implementation, while 33 (29.5\%) appear in the backend code. Figure~\ref{fig:complexconditional} presents two examples of \textit{Complex Conditional} from \href{https://github.com/Kashifraz/DIinAGP/blob/main/projects/P9_LMS/backend/src/main/java/com/lms/service/CourseContentService.java#L198}{\texttt{CourseContentService.java}} (\texttt{P9\_LMS}). The first conditional contains three conditional expressions, while the second contains four, which makes the code more difficult to understand. We focus on the second \textit{Complex Conditional}, which combines four conditions using three \texttt{OR} operators and one nested \texttt{AND} operator. This conditional performs content-type checks, filename extension checks, and null checks. Such conditionals introduce maintenance challenges because adding new functionality (e.g., supporting a new content type) typically requires additional conditional expressions, further increasing complexity (a violation of the \textit{KISS} principle). Similarly, \href{https://github.com/Kashifraz/DIinAGP/blob/main/projects/P1_CVbuilder/frontend/src/components/Awards/AwardFormModal.tsx#L145}{\texttt{AwardFormModal.tsx}} (\texttt{P1\_CVbuilder}) contains a \textit{Complex Conditional} with six conditional expressions linked by \texttt{AND} operators. This example uses hard-coded field names within the conditional expressions. This design issue has several consequences: (1) it makes debugging and fault localization more difficult, as it is harder to identify which specific condition failed, and (2) it increases the risk of introducing bugs or regressions when changes are made.

\textbf{(5)} \textit{Overall Code Complexity} \textbf{76 (5.6\%)}. This category of design issues indicates that a source file contains many conditional statements (e.g., \texttt{if}, \texttt{for}, \texttt{while}) throughout its implementation. These design issues are detected based on the \textit{cyclomatic complexity} metric. Unlike \textit{Complex Method}, this category focuses on the \textit{cyclomatic complexity} of the entire file.  Interestingly, code files with \textit{Overall Code Complexity} issues often exhibit additional design issues, such as \textit{Complex Methods}, \textit{Large Methods}, and \textit{Code Duplication} within a single file, indicating a significant design challenge. We quantitatively analyzed \textit{Overall Code Complexity} design issues and found that they occur more frequently in frontend code (47, 61.8\%) than in backend code (29, 38.1\%). As an example, we consider \href{https://github.com/Kashifraz/DIinAGP/blob/main/projects/P1_CVbuilder/backend/controllers/pdfController.js}{\texttt{PdfController.js}} (\texttt{P1\_CVBuilder}). This file contains 1,693 LoC. Within this file, three methods are identified as \textit{Complex Methods} (e.g., Line 6 and 108), seven as \textit{Large Methods} (e.g., Line 144 and 376), and three as \textit{Code Duplication} (e.g., Line 376 and 1197). (1) This code file exemplifies a violation of \textit{Separation of Concerns} (SoC) \cite{tarr1999ndegrees} design principle because the controller is implementing tasks it was not intended to handle (e.g., generating PDF templates from HTML and CSS). The controller (\href{https://github.com/Kashifraz/DIinAGP/blob/main/projects/P1_CVbuilder/backend/controllers/pdfController.js}{\texttt{PdfController.js}}) directly handles UI-related tasks for PDF generation, which increases the file's complexity and length. (2) Due to the PDF templates being hard-coded in JavaScript functions and the lack of a proper template system, excessive code duplication occurs. Each template function reinvents the wheel by reimplementing the HTML structure and CSS styles from scratch. This creates a maintainability challenge and also violates the \textit{DRY} design principle.

\begin{figure}[h]
    \centering
    \includegraphics[width=0.8 \linewidth]{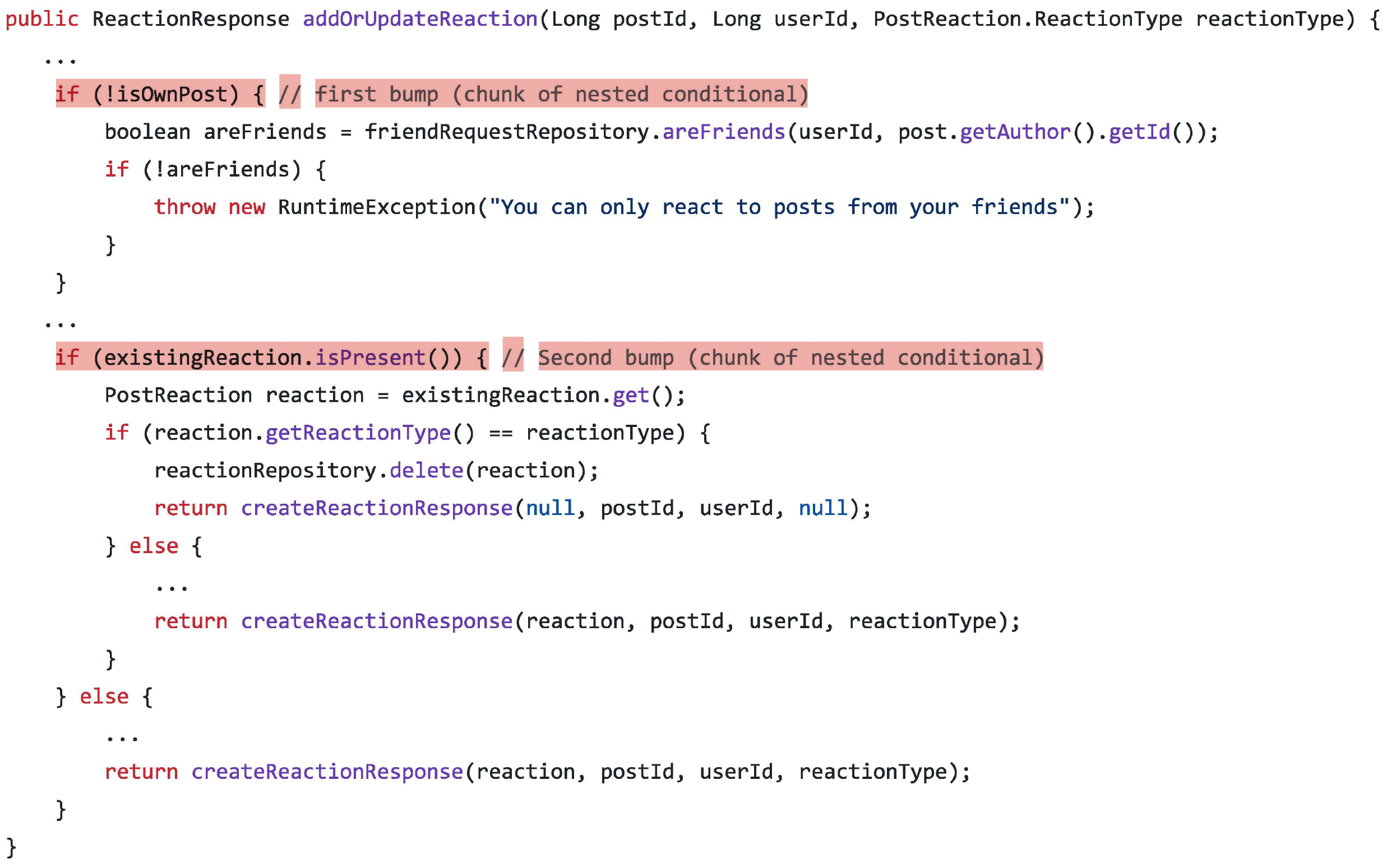}
    \caption{An example of \textit{Bumpy Road Ahead} design issue}
    \label{fig:bumpyroad}
    \vspace{-1em}
\end{figure}

\textbf{(6)} \textit{Bumpy Road Ahead} \textbf{79 (5.85\%)}. This category of design issues arises when a function contains multiple chunks of nested conditional logic. Such design issues reflect a lack of encapsulation of decision logic, which would otherwise help hide complexity and therefore become an obstacle to code comprehension. A quantitative analysis of \textit{Bumpy Road Ahead} design issues shows that they occur more frequently in backend code (48, 60.7\%) than in frontend code (31, 39.2\%). As an example, we consider the \texttt{addOrUpdateReaction()} method in \href{https://github.com/Kashifraz/DIinAGP/blob/main/projects/P8_SocialApp/backend/src/main/java/com/socialapp/service/ReactionService.java#L51}{\texttt{ReactionService.java}} (\texttt{P8\_SocialApp}). This method contains two chunks of nested conditional logic, which the CodeScene tool refers to as \textit{bumps} (see Figure \ref{fig:bumpyroad}). The first chunk of nested conditionals checks whether the user is authorized to react (e.g., \textit{like}) to a post. The second chunk checks whether the user has already reacted to the post. These nested chunks of conditional logic, illustrated in Figure~\ref{fig:bumpyroad}, increase developers' cognitive load when understanding the code (a violation of the \textit{KISS} principle), as they must track multiple execution paths and decision points to fully grasp it.

\begin{figure}[h]
    \centering
    \includegraphics[width=0.65 \linewidth]{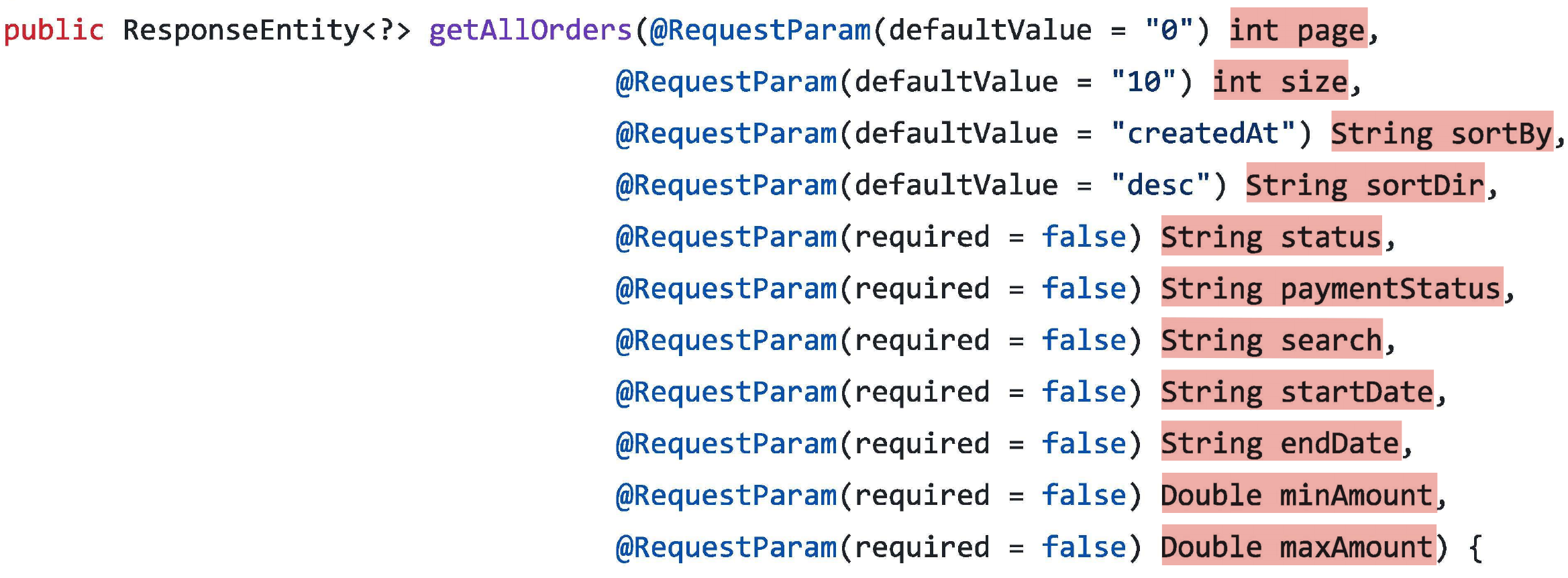}
    \caption{An example of \textit{Excess Number of Function Arguments} and \textit{Primitive Obsession} design issues}
    \label{fig:arguments}
    \vspace{-1em}
\end{figure}

\textbf{(7)} \textit{Primitive Obsession} \textbf{60 (4.44\%)}. These design issues indicate that functions in a code file contain too many primitive types (e.g., \texttt{int}, \texttt{double}, \texttt{float}) in their argument lists. It reflects a missing domain language (e.g., domain-specific types or classes), which signals a lack of abstraction in the code and introduces several challenges. First, primitive types require separate validation logic in the application code. Second, primitive types can lead to fragile code because they do not constrain the value range as a domain-specific type would. We found that \textit{Primitive Obsession} design issues occur more frequently in backend code (14, 23.3\%) than in frontend code (46, 76.7\%). For instance, \href{https://github.com/Kashifraz/DIinAGP/blob/main/projects/P3_Ecommerce/backend/src/main/java/com/ecommerce/controller/OrderController.java#L127}{\texttt{OrderController.java}} (\texttt{P3\_Ecommerce}) has several methods that use primitive types in their argument lists, which triggered the CodeScene tool to detect the \textit{Primitive Obsession} design issues. One example method from \href{https://github.com/Kashifraz/DIinAGP/blob/main/projects/P3_Ecommerce/backend/src/main/java/com/ecommerce/controller/OrderController.java#L127}{\texttt{OrderController.java}} is shown in Figure \ref{fig:arguments}, which has a long list of 11 arguments and also uses primitive types. 

\textbf{(8)} \textit{Excess Number of Function Arguments} \textbf{23 (1.7\%)}. These design issues arise when a function has an excessive number of arguments. Functions with too many arguments may indicate that the function handles multiple responsibilities, or a missing abstraction that could otherwise encapsulate these arguments. The \textit{Excess Number of Function Arguments} issues are found predominantly in backend code (18, 78.3\%) compared to frontend code (5, 21.7\%). As an example, we consider the \href{https://github.com/Kashifraz/DIinAGP/blob/main/projects/P3_Ecommerce/backend/src/main/java/com/ecommerce/controller/OrderController.java#L127}{\texttt{getAllOrders()}} method in the \texttt{OrderController.java} (\texttt{P3\_Ecommerce}). This method implements a RESTful API endpoint for retrieving all orders for the admin and includes 11 arguments in its parameter list, as also shown in Figure~\ref{fig:arguments}. The method attempts to handle multiple responsibilities, including pagination, sorting, filtering, and searching of orders (violation of \textit{SRP} principle). Such a design can break all callers when a parameter is changed, removed, or modified, making the order of parameters critical. This design also reduces code readability and complicates testing due to the large number of possible parameter combinations.

\textbf{(9)} \textit{Deep Nested Complexity} \textbf{15 (1.03\%)}. These design issues arise when code (e.g., \texttt{methods}) contains control structures (e.g., \texttt{if}-statements or loops) nested within other control structures. \textit{Deep Nested Complexity} increases developers' cognitive load when reading code. The \textit{Deep Nested Complexity} issues occur primarily in backend code (13, 86.7\%) rather than frontend code (2, 13.3\%). For instance, the \href{https://github.com/Kashifraz/DIinAGP/blob/main/projects/P6_FormPlugin/includes/field-types/class-dropdown-field.php\#L225}{\texttt{validate(\$value)}} method is used to validate whether the user-selected inputs are valid and handle single and multiple selections in a drop-down list. This method in the \texttt{P6\_FormPlugin} project uses control structures that are 4 levels deep, indicating a violation of the \textit{KISS} principle.  

\begin{tcolorbox}[
    colback=lightgray!20, 
    colframe=darkgray,   
    boxrule=0.5mm,        
    arc=2mm,              
    title=Finding 3,      
    fonttitle=\bfseries   
]
The design issues identified by CodeScene in AI IDE-generated projects include \textit{Code Duplication} and code complexity related issues such as \textit{Complex Methods}, \textit{Overall Code Complexity}, \textit{Bumpy Road Ahead}, \textit{Complex Conditionals}, and \textit{Large Methods}. These design issues indicate violations of the \textit{DRY}, \textit{SRP}, \textit{KISS}, and \textit{SoC} design principles, posing threats to the long-term maintainability and evolvability of the projects.  
\end{tcolorbox}

\subsubsection{RQ2.2: What design issues are identified by SonarQube in large-scale projects generated by Cursor?} \label{subsubsec:sonarqube}

\begin{figure}[h]
    \centering
    \includegraphics[width=1 \linewidth]{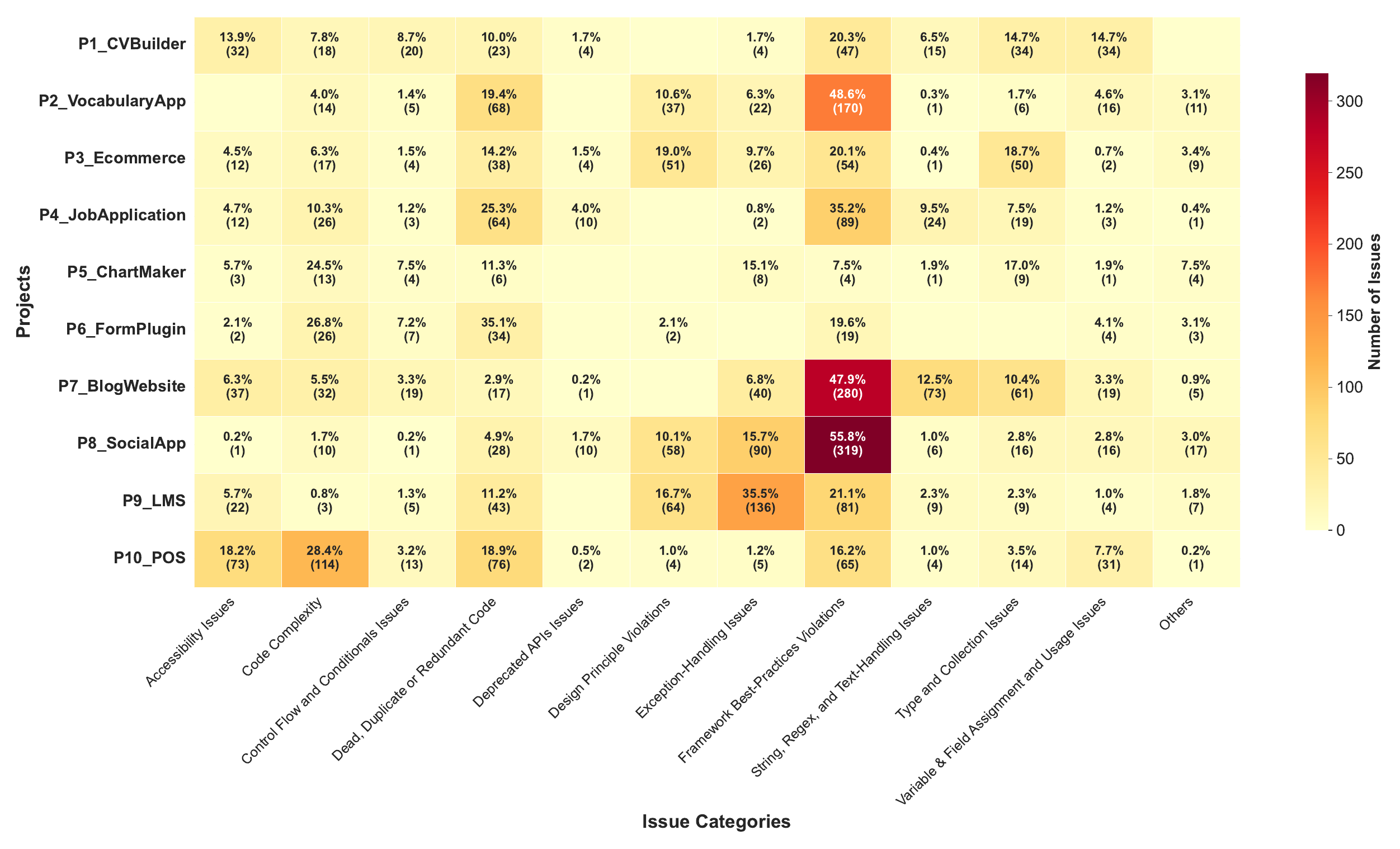}
    \caption{Overview of the design issues identified by SonarQube}
    \label{fig:sonarqube}
    \vspace{-1em}
\end{figure}

We also employed SonarQube to identify design issues in our Cursor-generated projects and identified 3,193 design issues categorized into 11 categories (see Figure \ref{fig:sonarqube}). The mapping of design issues identified by SonarQube to design principles is shown in Table \ref{tab:principle-mapping}.

\begin{table}[htbp]
\centering
\small
\caption{Overview of project issues by severity level}
\label{tab:severity}
\begin{tabular}{p{0.17\linewidth}p{0.1\linewidth}p{0.1\linewidth}p{0.08\linewidth}p{0.08\linewidth}p{0.06\linewidth}p{0.06\linewidth}}
\toprule
\textbf{Project} & \textbf{BLOCKER} & \textbf{CRITICAL} & \textbf{MAJOR} & \textbf{MINOR} & \textbf{INFO} & \textbf{Total} \\ \midrule
P1\_CVBuilder & 0 & 9 & 123 & 99 & 0 & 231 \\
P2\_VocabularyApp & 0 & 41 & 277 & 32 & 0 & 350 \\
P3\_Ecommerce & 8 & 65 & 145 & 48 & 2 & 268 \\
P4\_JobApplication & 0 & 2 & 154& 93 & 2 & 253\\
P5\_ChartMaker & 1 & 17 & 11 & 24 & 0 & 53 \\
P6\_FormPlugin & 0 & 31 & 50& 16& 0 & 97\\
P7\_BlogWebsite & 0 & 21 & 277& 286 & 0 & 584\\
P8\_SocialApp & 0 & 14 & 460 & 97 & 1 & 572 \\
P9\_LMS & 0 & 16 & 262 & 102 & 3 & 383 \\
P10\_POS & 0 & 93 & 211 & 97& 1 & 402\\
\midrule
\textbf{Total} & \textbf{9} & \textbf{309} & \textbf{1,970} & \textbf{894} & \textbf{9} & \textbf{3,193}\\ \bottomrule
\end{tabular}

\end{table}

Table \ref{tab:severity} presents the severity distribution of the design issues identified by SonarQube in Cursor-generated projects. The highest severity level is \textit{blocker}, followed by \textit{critical}, \textit{major}, \textit{minor}, and finally \textit{info}. The number of \textit{blocker} and \textit{info} issues is relatively small. However, there is a considerable number of \textit{critical} issues (309). The majority (1,970) of the identified issues fall under the \textit{major} category, and a substantial number (894) are classified as \textit{minor} issues. 

\textbf{(1)} \textit{Framework Best-Practice Violation \textbf{1,128 (35.3\%)}.} These issues reflect violations of recommended practices for the technologies being used (e.g., React prop validation, use of global objects). We illustrate an identified design issue ``\textit{`word' is missing in props validation}'' using an example from \href{https://github.com/Kashifraz/DIinAGP/blob/main/projects/P2_VocabularyApp/frontend/src/components/Flashcard.js#L14}{Flashcard.js} (\texttt{P2\_VocabularyApp}). This example shows that the ``word'' prop used in the component is not defined in the component’s PropTypes validation. The ``word'' prop has an expected structure and required fields, and it is recommended that React component props should be validated (e.g., via \texttt{PropTypes} or \texttt{TypeScript}) to: (1) prevent runtime errors if ``word'' is undefined or missing required fields, and (2) document the expected structure of the ``word'' object for other developers. We further present another issue, “\textit{Replace this use of System.out by a logger}”, using an example from \href{https://github.com/Kashifraz/DIinAGP/blob/main/projects/P2_VocabularyApp/backend/src/main/java/com/vocabularyapp/controller/QuizController.java#L36}{\texttt{QuizController.java}} (\texttt{P2\_VocabularyApp})
, which extensively uses \texttt{System.out.println()} statements. These statements are recommended to be replaced with a logging framework, such as \texttt{org.slf4j.Logger}. This is because: (1) \texttt{System.out} does not allow filtering logs by their severity, and (2) it cannot be properly configured, such as disabling or adjusting logging behavior in production environments.

\begin{figure}[h]
    \centering
    \includegraphics[width=0.25 \linewidth]{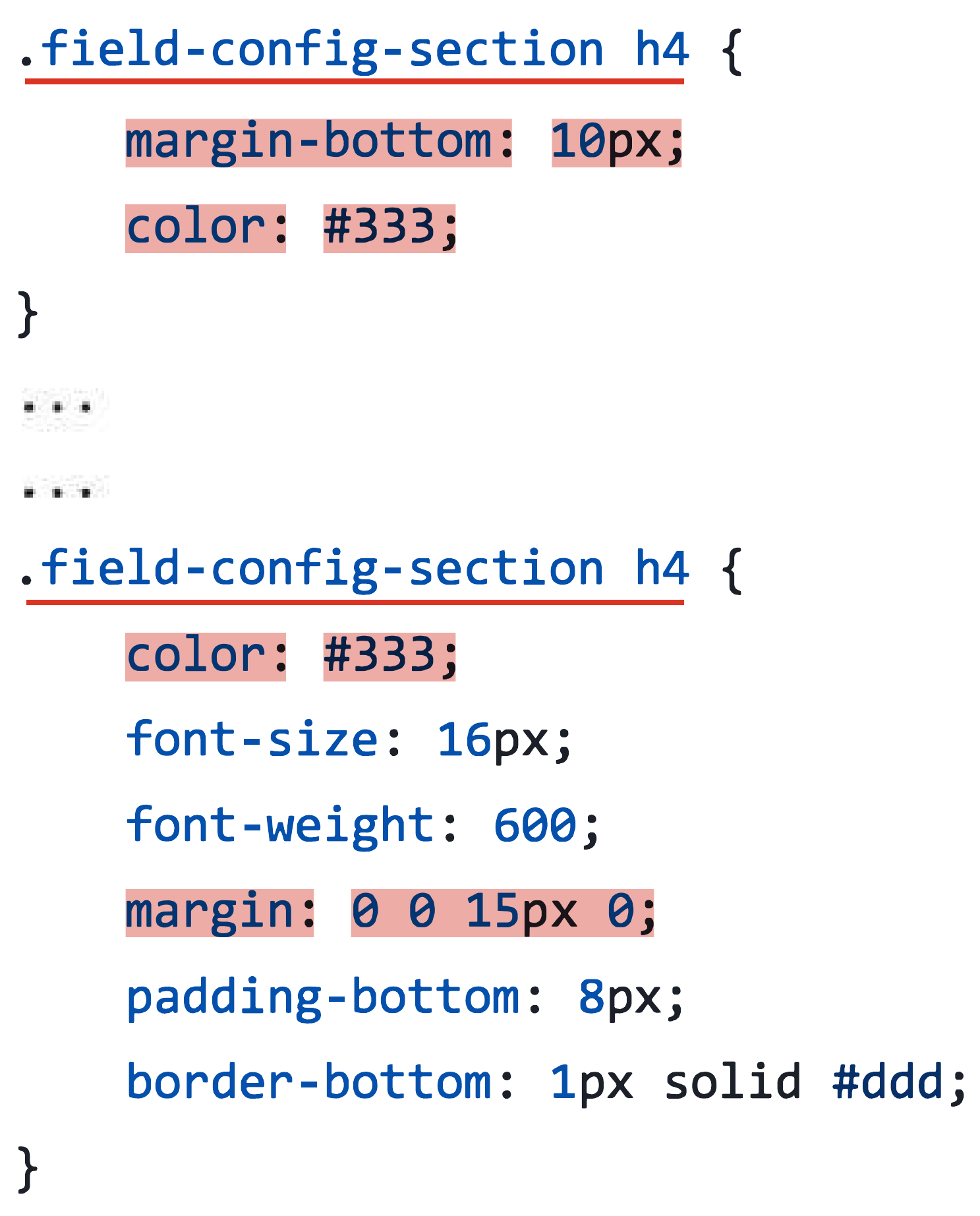}
    \caption{An example of \textit{Duplicated CSS Selectors} design issue}
    \label{fig:duplicate}
    \vspace{-1em}
\end{figure}

\textbf{(2)} \textit{Dead, Duplicate or Redundant Code \textbf{397 (12.4\%)}.} This category includes issues identified by SonarQube related to unused or duplicated code: unused imports or variables, duplicated string literals, and duplicate CSS selectors. These issues affect the readability of code and make the code harder to understand. For instance, we show an identified issue, ``\textit{Define a constant instead of duplicating this literal `Inventory item not found' 4 times}'', with an example from \href{https://github.com/Kashifraz/DIinAGP/blob/main/projects/P10_POS/backend/app/Controllers/Api/InventoryController.php#L364}{InventoryController.php} (\texttt{P10\_POS}) which duplicates the literal that is actually an error message 4 times. It should be replaced with a constant to avoid duplicating same literal 4 times to make code more maintainable. Similarly, we provide another issue, ``\textit{Unexpected duplicate selector `.field-config-section h4', first used at line 1075}'', with an example from \href{https://github.com/Kashifraz/DIinAGP/blob/main/projects/P6_FormPlugin/admin/css/admin.css#L1075}{admin.css} (\texttt{P6\_FormPlugin}) which contains duplicated CSS selectors (see Figure \ref{fig:duplicate}). The CSS selector `\texttt{.field-config-section h4}' is used twice in the same file with duplicate properties assignment, which is a maintainability challenge and indicate the violation of \textit{DRY} design principle.

\begin{figure}[h]
    \centering
    \includegraphics[width=0.85 \linewidth]{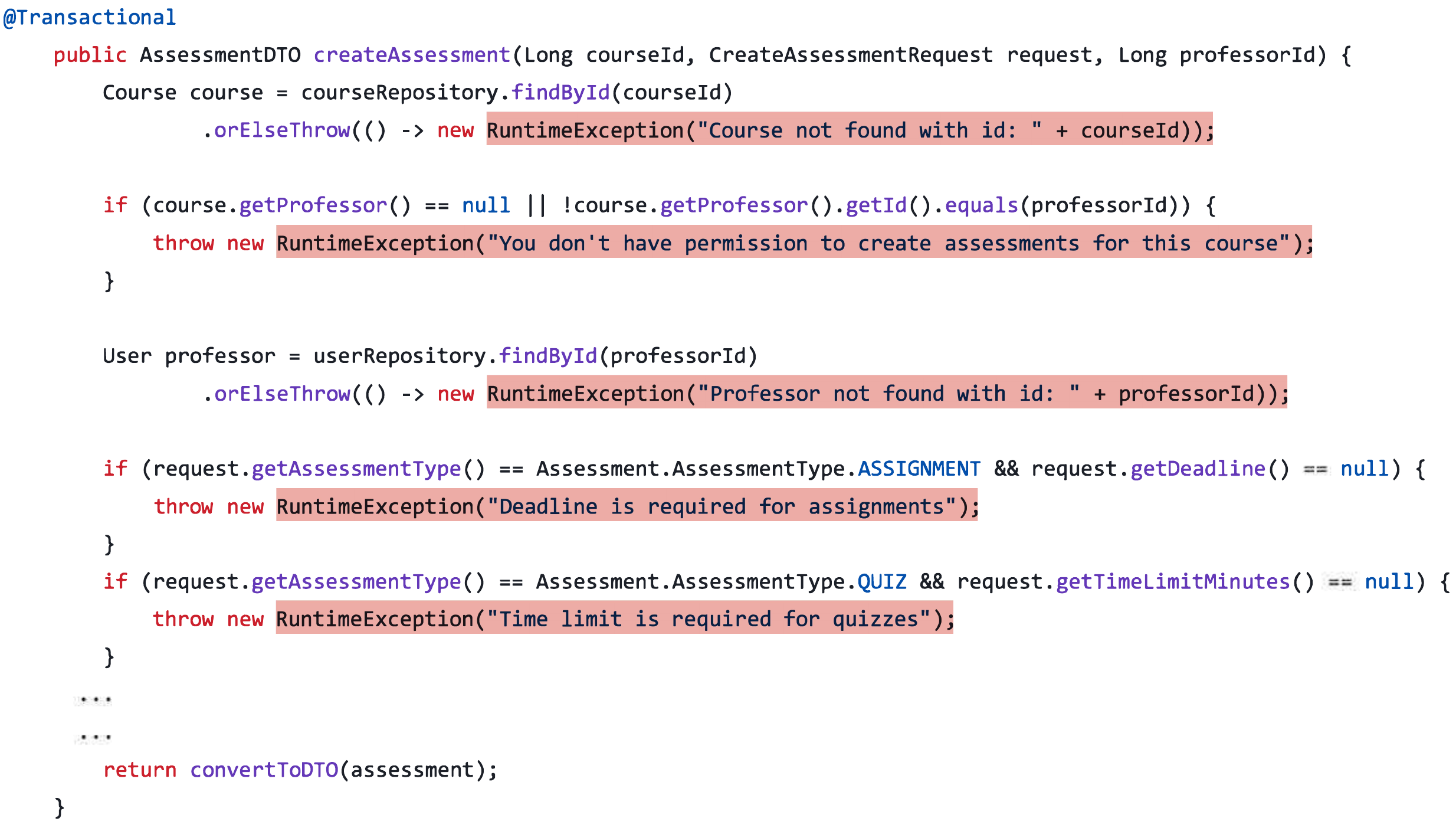}
    \caption{An example of \textit{Generic Exceptions} design issue}
    \label{fig:exception}
    \vspace{-1em}
\end{figure}

\textbf{(3)} \textit{Exception-Handling Issue \textbf{333 (10.4\%)}.} The issues in this category include inappropriate or incomplete exception handling such as catching overly generic exceptions, empty \texttt{catch} blocks, or failing to handle exceptions properly. We demonstrate an identified design issue, ``\textit{Replace generic exceptions with specific library exceptions or a custom exception}'', with an example from \href{https://github.com/Kashifraz/DIinAGP/blob/main/projects/P9_LMS/backend/src/main/java/com/lms/service/AssessmentService.java#L36}{\texttt{AssessmentService.java}} (\texttt{P9\_LMS}). This class performs improper exception handling by consistently throwing generic \texttt{RuntimeException} instances across all error scenarios (e.g., resource not found, permission violations, business rule validation failures) (see Figure \ref{fig:exception}). These generic exceptions do not provide sufficient information about the specific error type, preventing developers from accurately understanding the cause of the failure and implementing appropriate handling logic. Instead, custom exception classes or more specific Java exception types should be used. In the same manner, we present a design issue, ``\textit{Add logic to this catch clause or eliminate it and rethrow the exception automatically}'', with an example from \href{https://github.com/Kashifraz/DIinAGP/blob/main/projects/P8_SocialApp/frontend/src/services/commentService.js#L43}{\texttt{commentService.js}} (\texttt{P8\_SocialApp}). In this file, multiple asynchronous methods send HTTP requests to RESTful API endpoints for comment-related operations. However, these methods use redundant \textit{try-catch} blocks that do not properly handle errors (e.g., network or server errors) and simply rethrow the exceptions. This results in unnecessary code bloat without providing any meaningful error-handling logic. These issues also indicate the violation of \textit{Fail Fast} design principle \cite{shore2004fail}.

\begin{figure}[h]
    \centering
    \includegraphics[width=0.55 \linewidth]{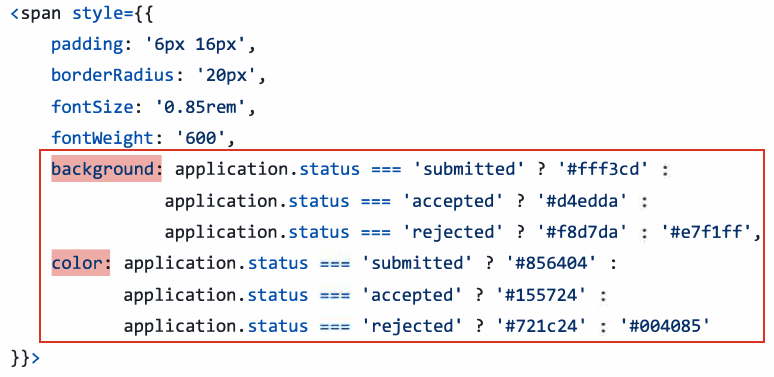}
    \caption{An example of \textit{Complex Ternary Operators} design issue}
    \label{fig:ternary}
    \vspace{-1em}
\end{figure}

\textbf{(4)} \textit{Code Complexity \textbf{273 (8.5\%)}.} This category focuses on issues related to complexity that make code difficult to understand and test, including high cognitive complexity and deeply nested expressions. We illustrate this category with an identified design issue, ``\textit{Refactor this function to reduce its Cognitive Complexity from 37 to the 15 allowed}'', in the \texttt{update} method from \href{https://github.com/Kashifraz/DIinAGP/blob/main/projects/P10_POS/backend/app/Controllers/Api/ProductController.php#L188}{ProductController.php} (\texttt{P10\_POS}) as an example. This method has a cognitive complexity of 37 due to excessive nested conditional logic and a large number of \texttt{if} statements. It should be simplified or split into smaller methods to reduce the cognitive complexity below 15. Similarly, we present another design issue, ``\textit{Extract this nested ternary operation into an independent statement}'', with an example from \href{https://github.com/Kashifraz/DIinAGP/blob/main/projects/P4_JobApplication/frontend/src/components/dashboard/EmployeeDashboard.js#L199}{\texttt{EmployeeDashboard.js}}
 (\texttt{P4\_JobApplication}) shown in Figure \ref{fig:ternary}. In this case, nested ternary operators are used to conditionally render the \texttt{background color} and \texttt{text color} of the application status, with three levels of nesting. This increases the cognitive complexity of the code and makes it harder to read and maintain. These issues also indicate a violation of the \textit{KISS} principle.


\begin{figure}[h]
    \centering
    \includegraphics[width=0.8 \linewidth]{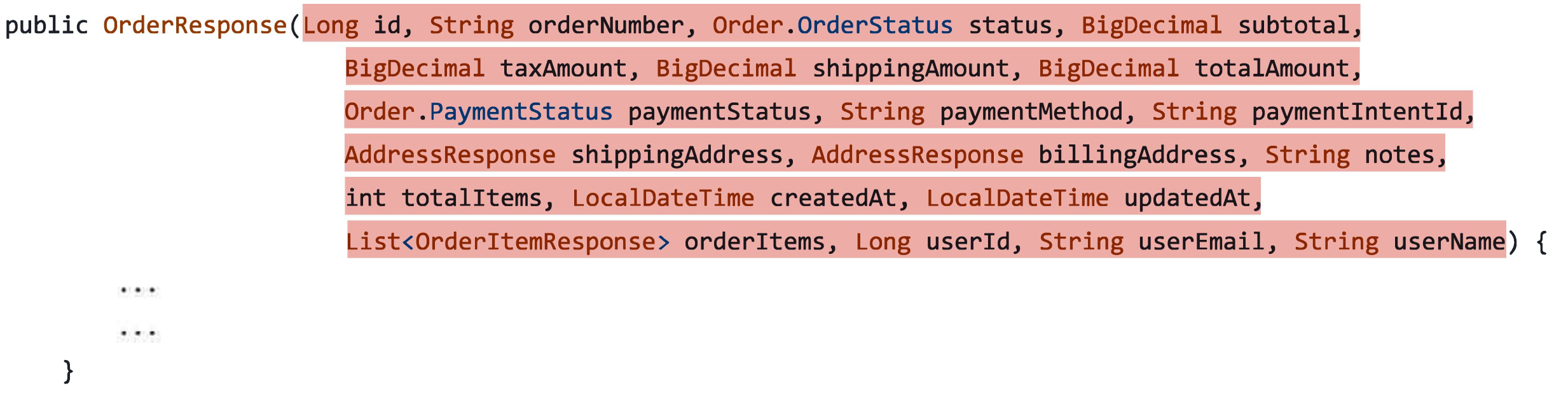}
    \caption{An example of a \textit{Higher Number of Constructor Parameters} design issue}
    \label{fig:design}
    \vspace{-1em}
\end{figure}

\textbf{(5)} \textit{Design Principle Violation \textbf{216 (6.8\%)}.} This category includes issues that violate design principles, such as keeping units of code (e.g., constructors, methods) small and focused. Similarly, we demonstrate an identified issue, ``\textit{Constructor has 20 parameters, which is greater than 7 authorized}'', with an example from \href{https://github.com/Kashifraz/DIinAGP/blob/main/projects/P3_Ecommerce/backend/src/main/java/com/ecommerce/dto/OrderResponse.java#L44}{\texttt{OrderResponse.java}} (\texttt{P3\_Ecommerce}) that contains a constructor with 20 parameters (see Figure \ref{fig:design}). This violates the SRP because this class handles orders, addresses, payments, and user details. We present another issue, ``\textit{This method has 11 returns, which is more than the 3 allowed}'', with an example from \href{https://github.com/Kashifraz/DIinAGP/blob/main/projects/P10_POS/backend/app/Controllers/Api/TransactionController.php#L712}{\texttt{TransactionController.php}} (\texttt{P10\_POS}). This example method has 11 exit points (e.g., \texttt{return} statements), making it hard to test. Multiple return statements often indicate that a method is performing too many unrelated tasks - a violation of the Single Responsibility Principle (\textit{SRP}).

\textbf{(6)} \textit{Type and Collection Issue \textbf{218 (6.8\%)}.} The issues in this category are related to type safety and the appropriate use of collections and iteration (e.g., use of `\texttt{for…of}', keys in lists). For instance, we demonstrate an identified issue, ``\textit{Use `\texttt{for…of}' instead of `\texttt{.forEach(…)}'}'', using an example from \href{https://github.com/Kashifraz/DIinAGP/blob/main/projects/P1_CVbuilder/backend/controllers/educationController.js#L81}{\texttt{educationController.js}} (\texttt{P1\_CVbuilder}). This issue suggests preferring `\texttt{for…of}' when iterating over collections because it provides better control flow, as it supports \texttt{break}, \texttt{continue}, and \texttt{return} statements, which \texttt{.forEach()} does not allow. Additionally, it works more effectively with \texttt{async/await}. We also illustrate an issue, ``\textit{Do not use Array index in keys}'', with an example from \href{https://github.com/Kashifraz/DIinAGP/blob/main/projects/P4_JobApplication/frontend/src/components/jobs/JobDetail.js#L200}{\texttt{JobDetail.js}} (\texttt{P4\_JobApplication}). This issue occurs due to the use of array indices as React keys in multiple places when rendering lists, which can cause performance issues. React may not properly identify which items have changed, been added, or removed, leading to unnecessary re-renders. It can also result in incorrect list rendering if the order changes or items are inserted or deleted.

\begin{figure}[h]
    \centering
    \begin{minipage}[b]{0.4\textwidth}
        \centering
        \includegraphics[width=\linewidth]{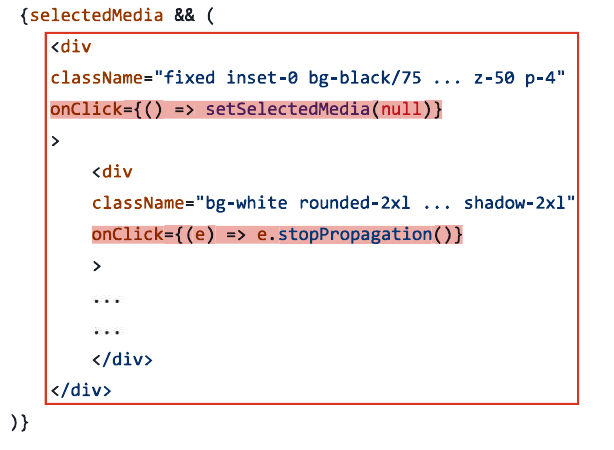}
        \caption{An example of a \textit{Non-Native Interactive Element} design issue}
        \label{fig:accessibility}
    \end{minipage}
    \hfill
    \begin{minipage}[b]{0.5\textwidth}
        \centering
        \includegraphics[width=\linewidth]{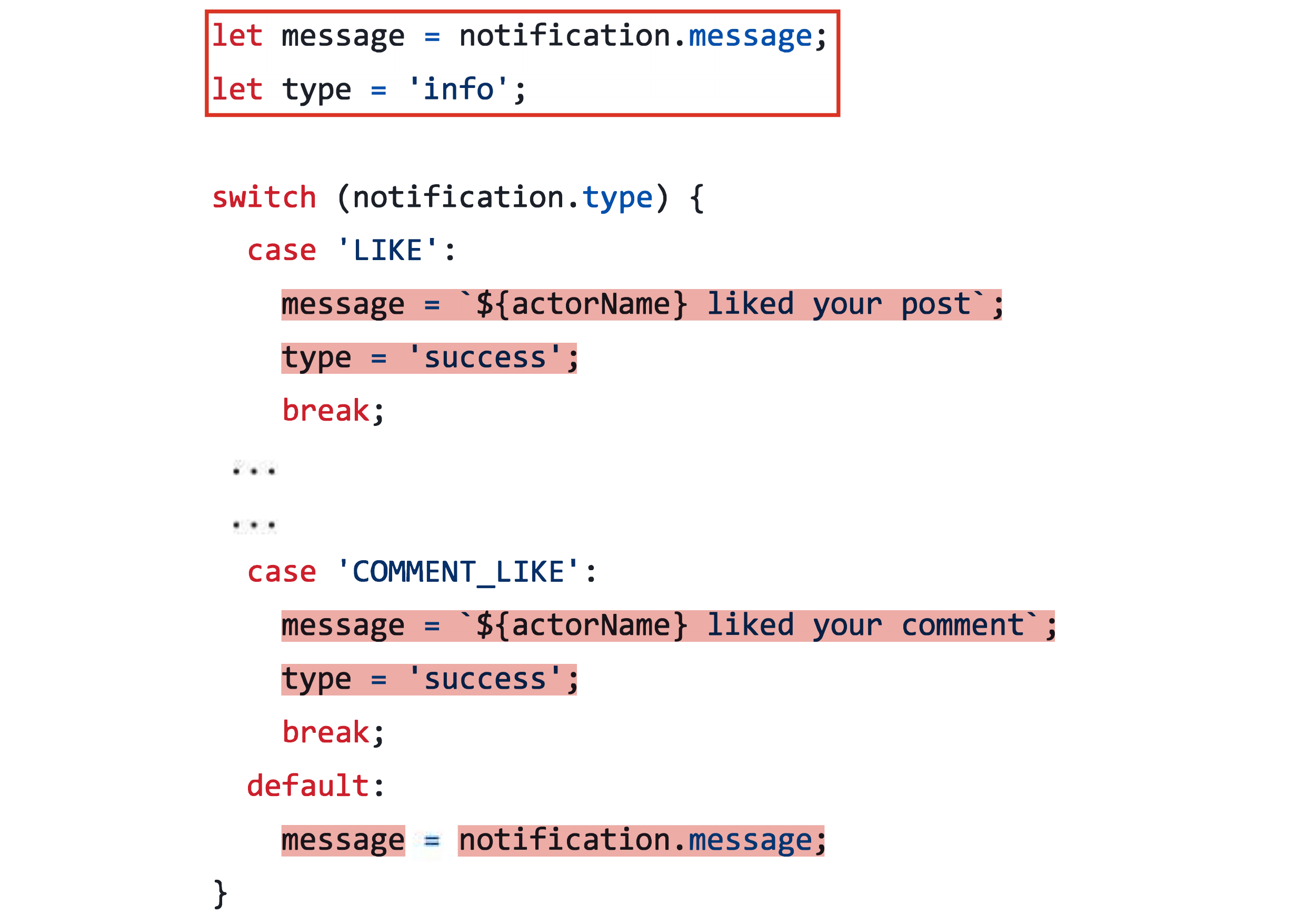}
        \caption{An example of \textit{Useless Assignment to a Variable} design issue}
        \label{fig:assignment}
    \end{minipage}
    \vspace{-1em}
\end{figure}

\textbf{(7)} \textit{Accessibility Issue \textbf{194 (6.1\%)}.}  This category flags missing or improper associations between labels and form controls, non-accessible interactive elements, and invalid or non-navigable link targets. These issues hinder accessibility across different devices. For instance, we illustrate an identified design issue, ``\textit{A form label must be associated with a control}'', with an example from \href{https://github.com/Kashifraz/DIinAGP/blob/main/projects/P7_BlogWebsite/admin_frontend/src/components/CarouselBlock.jsx#L267}{\texttt{CarouselBlock.jsx}} (\texttt{P7\_BlogWebsite}). The label \textit{image URL} is not associated with the \textit{image URL} input field.  This issue limits accessibility for other devices such as screen readers and these labels should be correctly linked to input fields. Similarly, we present another accessibility issue shown in Figure \ref{fig:accessibility}, ``\textit{Avoid non-native interactive elements. If using native HTML is not possible, add an appropriate role and support for tabbing, mouse, keyboard, and touch inputs to an interactive content element}''. We demonstrate this issue using an example from \href{https://github.com/Kashifraz/DIinAGP/blob/main/projects/P7_BlogWebsite/admin_frontend/src/pages/MediaLibraryPage.jsx#L366}{\texttt{MediaLibraryPage.jsx}} (\texttt{P7\_BlogWebsite}) which uses \texttt{onclick()} event handler on a non-native interactive element \texttt{<div>}. In addition, it only defines an \texttt{onClick} event handler and does not provide equivalent support for keyboard interactions (e.g., \texttt{onKeyDown}, \texttt{onKeyPress}) or proper accessibility roles. This issue represents a fundamental accessibility violation that creates barriers for users relying on assistive technologies, keyboard navigation, or alternative input methods.

\textbf{(8)} \textit{Variable and Field Assignment \& Usage Issue \textbf{130 (4.1\%)}.} These issues include incorrect use of variables and declarations (e.g., lexical declarations in \texttt{switch} cases, use of `var'). We demonstrate an identified design issue (see Figure \ref{fig:assignment}) with the example from \href{https://github.com/Kashifraz/DIinAGP/blob/main/projects/P8_SocialApp/frontend/src/context/NotificationContext.js#L39}{\texttt{NotificationContext.js}} (\texttt{P8\_SocialApp}) suggesting that, ``\textit{Remove this useless assignment to variable `message'}''. The statement \texttt{let message = notification.message;} declares \texttt{message} variable and performs a useless assignment to the variable. This is because the \texttt{message} variable is overwritten in all explicit cases of the \texttt{switch} statement. A better approach would be to declare the variable without initialization (e.g., \texttt{let message;}). Similarly, we present another example, ``\textit{Unexpected lexical declaration in case block}'', which suggests that \texttt{let} or \texttt{const} in \texttt{switch} cases should be wrapped in a block to avoid scope and reuse errors. We demonstrate this issue with the example from \href{https://github.com/Kashifraz/DIinAGP/blob/main/projects/P7_BlogWebsite/admin_frontend/src/pages/PostPreviewPage.jsx#L342}{\texttt{PostPreviewPage.jsx}} (\texttt{P7\_BlogWebsite}). In this example, the variables \texttt{HeadingTag} and \texttt{headingClasses} are declared inside the \texttt{heading} case without being wrapped inside curly braces. Without the block, these \texttt{const} declarations are scoped to the entire \texttt{switch} statement.

\begin{figure}[h]
    \centering
    \includegraphics[width=0.65 \linewidth]{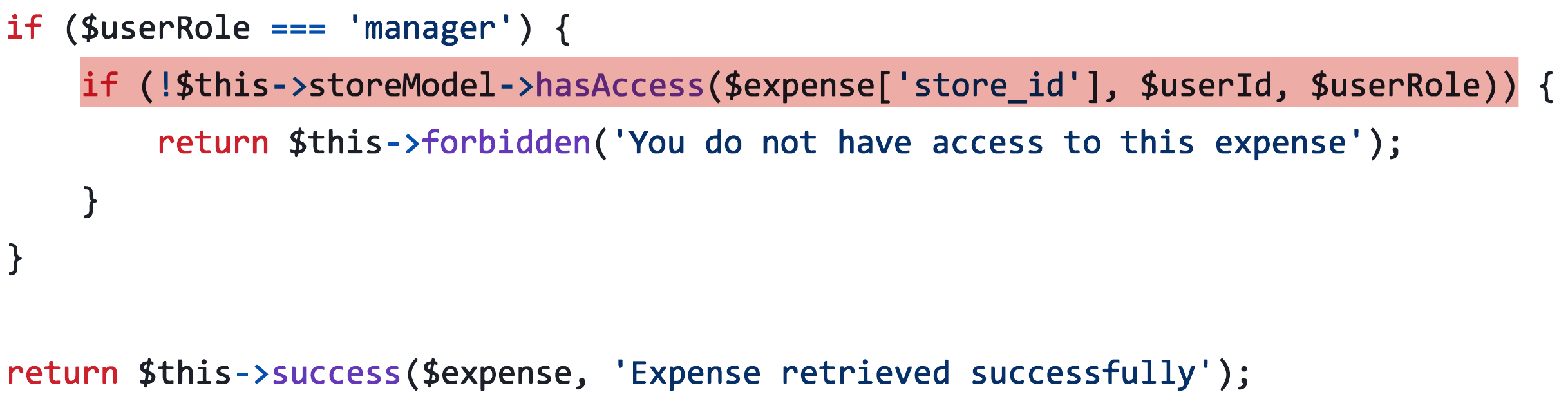}
    \caption{An example of \textit{Merging Conditional} design issue}
    \label{fig:conditional}
    \vspace{-1em}
\end{figure}

\textbf{(9)} \textit{Control Flow and Conditionals Issue \textbf{81 (2.5\%)}.} These issues are related to clarity and complexity of conditional logic and control flow (e.g., negated conditions, merging conditions, use of booleans). We demonstrate an identified design issue, ``\textit{Unexpected negated condition}'', with an example from \href{https://github.com/Kashifraz/DIinAGP/blob/main/projects/P1_CVbuilder/backend/controllers/skillController.js#L106}{\texttt{SkillController.js}} (\texttt{P1\_CVbuilder}). This controller uses multiple negative conditional logics which are hard to read and it is suggested that conditions should be written in positive form to improve readability. Similarly, we present a design issue shown in Figure \ref{fig:conditional}, ``\textit{Merge this if statement with the enclosing one}'', from \href{https://github.com/Kashifraz/DIinAGP/blob/main/projects/P10_POS/backend/app/Controllers/Api/ExpenseController.php#L147}{ExpenseController.php} (\texttt{P10\_POS}), which contains nested \texttt{if} statements that can be merged. These nested conditionals unnecessarily increase the level of nesting and cognitive complexity (a violation of the \textit{KISS} principle). These nested conditionals should be combined into a single, clear condition to improve readability and maintainability.

\textbf{(10)} \textit{String, Regex, and Text-Handling Issue \textbf{134 (4.2\%)}.} This category includes issues related to best practices for handling strings, regular expressions, and escaping special characters. We present an example issue from \href{https://github.com/Kashifraz/DIinAGP/blob/main/projects/P4_JobApplication/frontend/src/components/profile/Profile.js\#L209}{\texttt{Profile.js}} (\texttt{P4\_JobApplication}) which states, ``\textit{HTML entity, `\}', must be escaped}''. The issue is triggered due to the use of the HTML entity ``\texttt{$\surd$}'' in the code \texttt{\{user?.verificationBadge \&\& <span className="verified-check">$\surd$</span>\}} instead of using the unicode character \texttt{\textbackslash{}u2713}. Similarly, we present another issue related to regular expressions from \href{https://github.com/Kashifraz/DIinAGP/blob/main/projects/P7_BlogWebsite/admin_frontend/src/components/TableOfContentsBlock.jsx#L25}{\texttt{TableOfContentsBlock.jsx}} (\texttt{P7\_BlogWebsite}) which states, ``\textit{Group parts of the regex together to make the intended operator precedence explicit}''. This issue is flagged due to ambiguous operator precedence in the regular expression passed to the \texttt{replace()} method, which removes leading or trailing hyphens to generate URL-friendly IDs. 

\textbf{(11)} \textit{Deprecated APIs Issue \textbf{31 (1.0\%)}.} These issues identify the use of deprecated language or library APIs that may break in future versions or have safer alternatives. For instance, we present an identified issue, ``\textit{The signature `(commandId: string): boolean' of `document.queryCommandState' is deprecated}'', with an example from \href{https://github.com/Kashifraz/DIinAGP/blob/main/projects/P8_SocialApp/frontend/src/components/RichTextEditor.web.js#L78}{\texttt{RichTextEditor.web.js}} (\texttt{P8\_SocialApp}). This issue is triggered by the \texttt{queryCommandState()} API, which has been officially deprecated by major browsers and may be removed in future versions. Similarly, we present another issue, ``\textit{The signature `(from: number, length?: number | undefined): string' of `text.substr' is deprecated}'', from an example of \href{https://github.com/Kashifraz/DIinAGP/blob/main/projects/P1_CVbuilder/frontend/src/utils/index.ts#L151}{index.ts} (\texttt{P1\_CVbuilder}). This issue is triggered by the use of the deprecated \texttt{substr()} API in multiple functions.

\begin{tcolorbox}[
    colback=lightgray!20, 
    colframe=darkgray,   
    boxrule=0.5mm,        
    arc=2mm,              
    title=Finding 4,      
    fonttitle=\bfseries   
]
The issues identified by SonarQube in AI IDE-generated projects include \textit{Framework Best-Practice Violations}, \textit{Dead, Duplicate or Redundant Code}, \textit{Exception-Handling Issues}, and \textit{Code Complexity}, among others. These design issues also indicate violations of the \textit{DRY}, \textit{SRP}, \textit{Fail Fast} and \textit{KISS} design principles, posing threats to the long-term maintainability and evolvability of the projects.
\end{tcolorbox}

\subsubsection{RQ2.3: What are the key overlapping design issues identified by both CodeScene and SonarQube?}  
\begin{figure}[h]
    \centering
    \includegraphics[width=0.6 \linewidth]{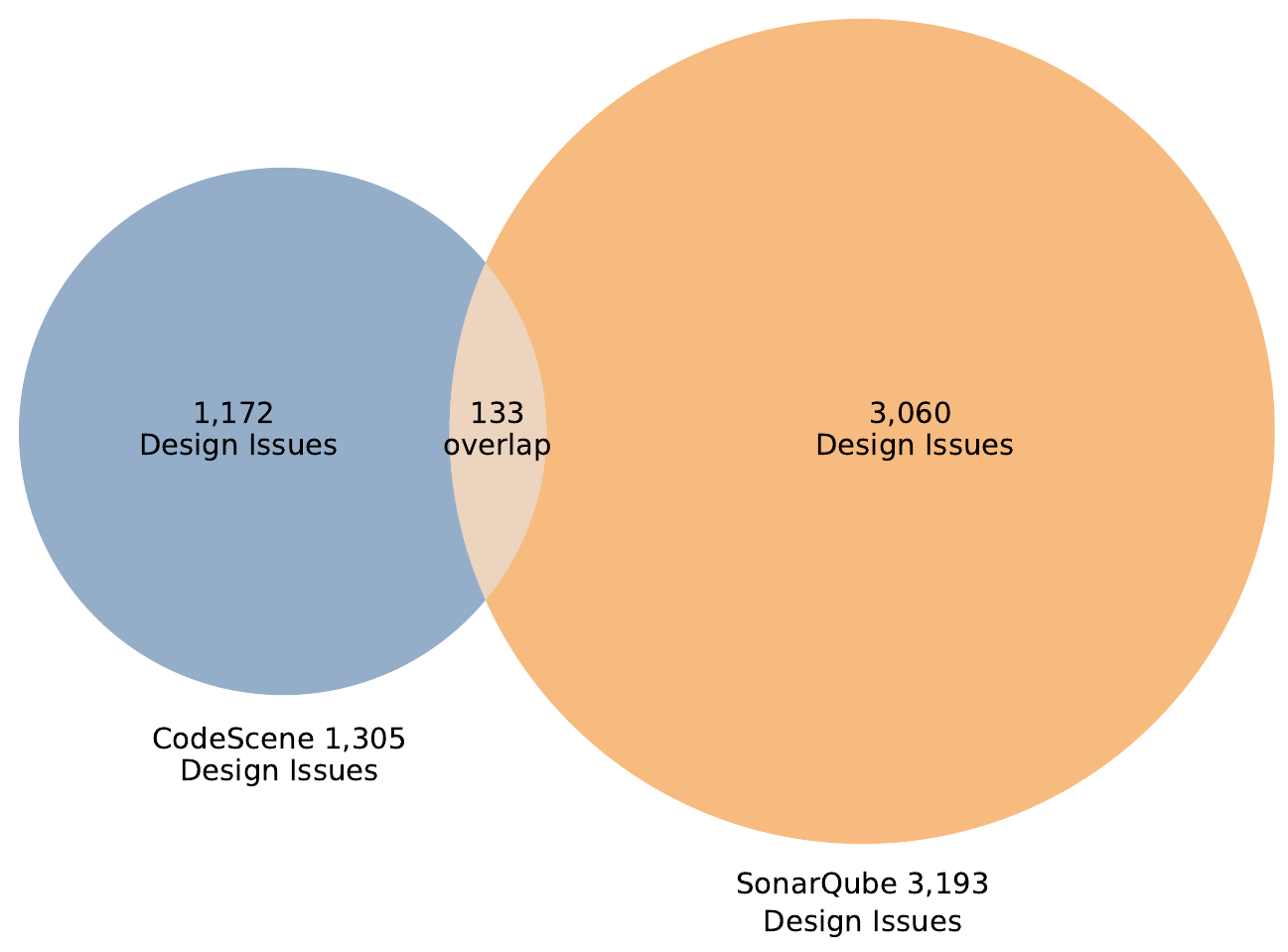}
    \caption{Overview of overlap between issues identified by SonarQube and CodeScene}
    \label{fig:overlap}
    \vspace{-1em}
\end{figure}

We employed two tools to identify design issues in 10 Cursor-generated projects. These two tools detect design issues at different levels of granularity. For example, CodeScene focuses more on high-level design issues (e.g., class and method-level issues), whereas SonarQube also detects design issues at a lower level (e.g., conditional and statement-level issues). However, we still found 133 overlapping design issues that were detected by both tools. In total, three issue patterns listed below, identified by SonarQube, overlap with \textit{Complex Method} and \textit{Large Method} issues identified by CodeScene. Issue patterns are the consolidated representation of recurring design issues identified by SonarQube (see Section \ref{subsubsec:sonarmethod}). Most of these overlapping issues are related to code complexity. Although the number of overlaps is not large, all the overlapping issues have \textit{critical} severity according to SonarQube. The amount of overlapping issues is shown in Figure \ref{fig:overlap}.

\textbf{(1)} ``\textit{Refactor this function to reduce its Cognitive Complexity from ... to the 15 allowed}.'' This issue pattern is identified by SonarQube 82 times across the 10 projects and categorized under the \textit{Code Complexity} category. SonarQube identifies these design issues based on the cognitive complexity metric. This issue pattern has a 100\% overlap with the \textit{Complex Method} issues, which were identified 377 times across the 10 projects by CodeScene. CodeScene identifies these \textit{Complex Method} issues based on the cyclomatic complexity metric. This means that all 82 cognitive complexity issues identified by SonarQube are also classified as \textit{Complex Methods} by CodeScene.

\textbf{(2)} ``\textit{This method has ... returns, which is more than the 3 allowed}.'' This is another issue pattern identified by SonarQube, found 100 times across the 10 projects. This design issue is triggered by SonarQube when a function has more than three exit points and categorized under the \textit{Code Complexity} category. These issue pattern have a 48\% overlap with the \textit{Complex Method} issues identified 377 times by CodeScene. This means that 48 out of the 100 methods with multiple exit points identified by SonarQube are also identified as \textit{Complex Methods} by CodeScene. This is because a higher number of \texttt{return} statements or exit points is also likely to increase the cyclomatic complexity of the methods.

\textbf{(3)} ``\textit{This function `...' has `...' lines, which is greater than the 150 lines authorized}.'' This issue pattern is identified only 3 times by SonarQube across the 10 projects and categorized under the \textit{Design Principle Violations} category. These issues overlap with the \textit{Large Method} issue pattern identified 171 times by CodeScene. The reason for the low overlap is that the default threshold for SonarQube is 150 LOC, and SonarQube is also less likely to detect function length issues in frontend code.

\begin{tcolorbox}[
    colback=lightgray!20, 
    colframe=darkgray,   
    boxrule=0.5mm,        
    arc=2mm,              
    title=Finding 5,      
    fonttitle=\bfseries   
]
The overlap between the design issues identified by CodeScene and SonarQube is low (133 design issues). All of these overlapping issues have \textit{Critical} severity according to SonarQube and most of them are related to code complexity. 
\end{tcolorbox}

\subsubsection{RQ2.4: What technology-specific design issues does Cursor consistently generate in large-scale projects?}  
The design issues detected by SonarQube consist of 1,900 (59\%) technology-specific issues across 38 unique issue patterns. These design issues are detected at the statement-level. In contrast, the design issues detected by CodeScene are general and not specific to any technology. We group these issues based on specific technologies in which they occur. Specifically, we present four categories of technology-specific issues: \textit{JavaScript/React/Node.js}, \textit{Java/Spring Boot}, \textit{HTML/CSS/JSX}, and \textit{PHP}.

\textbf{(1)} \textit{JavaScript/React/Node.js \textbf{1,216 (64\%)}.} SonarQube identified a significant number of issues related to JavaScript, including issues in vanilla JavaScript, React, and Node.js. For instance, we present a vanilla JavaScript issue pattern, ``\textit{Prefer `globalThis' over `window'}'', which is found 128 times across all projects. This issue suggests using the standardized \texttt{globalThis} keyword to access global variables instead of environment-specific keywords such as \texttt{window}, \texttt{global}, or \texttt{self}. We also present an issue pattern specific to React, ``\textit{`...' is missing in props validation}'', which is identified 540 times. This issue occurs when a prop used in a React component is not defined in PropTypes, preventing type checking and validation. Similarly, we present another Node.js specific issue pattern, ``\textit{Use the node: protocol for Node.js core module import}'', which is found 27 times across all projects. This issue suggests using the \texttt{node:} protocol to explicitly import Node.js core modules (e.g., \texttt{fs}, \texttt{path}), thereby avoiding conflicts with third-party packages that may have the same names.

\textbf{(2)} \textit{Java/Spring Boot \textbf{363 (19\%)}.} SonarQube also identified a considerable number of issues related to Java and Spring Boot. We present an issue pattern specific to Java, ``\textit{Replace `Stream.collect(Collectors.toList())' with `Stream.toList()'}'', which was identified 66 times. This issue suggests using the newer \texttt{Stream.toList()} method (Java 16+) instead of the older \texttt{Stream.collect(Collectors.toList())} pattern. Similarly, we present an issue pattern specific to Spring Boot, ``\textit{Remove this field injection and use constructor injection instead}'', which was found 187 times. This issue occurs when dependencies are injected directly into a class field (e.g., using \texttt{@Autowired}) rather than through the constructor, which reduces testability and hides class dependencies.

\textbf{(3)} \textit{HTML/CSS/JSX \textbf{285 (15\%)}.} SonarQube also detected issues related to HTML markup, including Vue.js HTML templates and JSX in frontend code. For example, we present a CSS-related issue pattern, ``\textit{Unexpected duplicate selector `...', first used at line `...'}'', which is detected 64 times. This issue occurs when CSS files contain duplicated selectors, which may lead to conflicting property definitions. In addition, we present a JSX-specific issue pattern, ``\textit{onMouseOver/onMouseOut must be accompanied by onFocus/onBlur}'', which appears 8 times. This is an accessibility issue triggered when mouse event handlers (e.g., \texttt{onMouseOver}, \texttt{onMouseOut}) are used without their keyboard-equivalent event handlers (e.g., \texttt{onFocus}, \texttt{onBlur}), making interactive elements inaccessible to keyboard users.

\textbf{(4)} \textit{PHP \textbf{33 (1.73\%)}.} SonarQube also identified a small number of issues specific to PHP. For instance, the issue pattern ``\textit{Replace `...' with namespace import mechanism through the `\texttt{use}' keyword}'' is found 22 times. This issue occurs when a class or function is referenced using its full namespace path instead of importing it into the current scope using the \texttt{use} keyword, which improves code readability and maintainability. Similarly, we present the issue patterns, ``\textit{Replace `\texttt{require}' with `\texttt{require\_once}'}'' and ``\textit{Replace `\texttt{include}' with `\texttt{include\_once}'}'', which together are found 8 times. These issues occur when files are included using \texttt{include} or \texttt{require} instead of \texttt{include\_once} or \texttt{require\_once}, which may lead to redeclaration errors if the same file is loaded multiple times.

\begin{tcolorbox}[
    colback=lightgray!20, 
    colframe=darkgray,   
    boxrule=0.5mm,        
    arc=2mm,              
    title=Finding 6,      
    fonttitle=\bfseries   
]
Issues identified by SonarQube are 59\% technology-specific. Majority (64\%) of these issues are related to \textit{JavaScript/React/Node.js}, followed by \textit{Java/Spring Boot} (19\%) and \textit{HTML/CSS/JSX} (15\%).
\end{tcolorbox}

\section{Discussion} \label{sec:discussion}

\subsection{Building Large-Scale Projects using AI IDEs.}

\subsubsection{Human Efforts Concentrated More on the Higher-Level Tasks.} Our results show that successful large-scale project generation with Cursor appears to depend more on investing human effort in higher-level tasks, such as requirements and design-level activities. Human involvement in assisting AI IDEs to produce clear functional requirements, architectural decisions, and a decomposition of the system into independently testable features with explicit acceptance criteria improves the functional correctness of the projects. This is because, once these higher-level artifacts are in place, low-level implementation tasks can be delegated to the AI IDE with human oversight, which accelerates development velocity while keeping project quality under human control. A similar perspective is also reflected in OpenAI’s recent \textit{Harness Engineering} view in agent-first development: \textit{humans steer, agents execute} \cite{lopopolo2025harness}. This aligns with established incremental delivery models in software engineering (e.g., feature-driven development \cite{palmer2001practical}) and is consistent with recent qualitative evidence that requirements, as typically documented, are often too abstract to be used directly as LLM inputs. Instead, requirements need to be decomposed into programming tasks and enriched with design and architectural constraints before prompting \cite{ullrich2025requirements}. 

\subsubsection{Ad Hoc Prompting vs. Systematic Approach.} Becker \textit{et al.} reported that AI IDE assistance did not improve productivity for experienced developers in complex open-source tasks, but instead slowed their performance \cite{becker2025measuring}. One possible explanation is that developers may not have used a systematic approach, which limited the effectiveness of AI IDEs. 
As we also found in our pilot study, approaching a complex task without decomposing it into sub-tasks and relying on an ad hoc (non-systematic and unplanned) approach (e.g., \textit{Vibe Coding} \cite{fawzy2025vibe}) can result in the AI IDEs leading the task with limited human control, diminishing the AI IDEs' effectiveness. Hence, there is a need to augment human intent in the development process by enabling humans to drive development while leveraging the full code generation potential of AI IDEs. Our study applies a systematic Feature-Driven Human-In-The-Loop (FD-HITL) framework rather than ad hoc prompting to generate projects. 
Our results suggest that such a structured workflow can better support complex development tasks. 


\subsection{Design Issues Identified in AI IDE-generated projects.}

\subsubsection{The Most Common Design Issues Identified in Cursor-Generated Projects.} 
The comparison of our results to existing literature shows similar or complementary findings; for example, Huang \textit{et al.} found that LLM agents frequently disregard opportunities for code reuse, resulting in greater redundancy than human developers \cite{huang2026more}. These results are consistent with our findings, in which we identified \textit{Code Duplication} as the most frequent design issue (28.4\% of issues identified by CodeScene). Similarly, Cotroneo \textit{et al.} \cite{cotroneo2025human} report that LLM-generated code tends to be simpler and more repetitive than human-written code in their large-scale comparison of code authored by human developers and LLMs. Our results present a complementary, but not identical, picture at the project level: Cursor-generated projects frequently exhibit substantial method-level and file-level complexity. Our study identified \textit{Overall Code Complexity}, \textit{Complex Method}, and \textit{Complex Conditional} design issues using the CodeScene tool, and \textit{Code Complexity} issues using the SonarQube tool. We attribute this difference partly to the shift from LLM-generated code snippets to end-to-end project generation using AI IDEs with agentic capabilities (e.g., autonomous planning and interaction with the codebase). These results are also aligned with He \textit{et al.} who found that the adoption of Cursor leads to an increase in project-level development velocity but also increases static analysis issues and code complexity \cite{he2025speed}. Interestingly, prior work shows that \textit{Code Duplication} and \textit{Exception-Handling Issues} are the most common and costly issues in human-developed systems \cite{digkas2017evolution}. Our Cursor-generated projects also exhibit \textit{Code Duplication} and \textit{Exception-Handling Issues} along with other frequently occurring issues including \textit{Framework Best-Practice Violations}, high \textit{Code Complexity}, \textit{Large Methods} and \textit{Accessibility Issues}.


These results indicate that although AI IDEs can produce large-scale projects with high functional correctness when operated systematically, they still exhibit a large number of design issues. This suggests that AI IDE-generated projects may work as prototypes but may not meet the quality standards needed for industrial adoption at scale due to long-term maintainability and evolvability challenges, requiring close supervision by experienced developers. This implies that AI IDEs require continuous human judgment and, as currently stands, may not immediately replace developers' engineering skills, especially for high-level tasks such as refining requirements and decomposing the project into independently testable features.

\subsubsection{Correlation and Overlap in Design Issues across Projects.} Design issues detected by CodeScene are not isolated but found to be correlated with other issues. For instance, files detected with \textit{Overall Code Complexity} also exhibit other design issues such as \textit{Complex Method}, \textit{Large Method}, and \textit{Code Duplication}, which usually make the overall code file complex. Similarly, in \textit{Large Methods}, it is also likely to identify \textit{Complex Methods}, because excessive code length can increase code complexity. Prior empirical studies also observed that size-related and complexity-related metrics are often positively correlated in practice \cite{mamun2017correlations}. In addition, the issues identified by CodeScene and SonarQube also overlap on a relatively small subset (133) of design issues (see Figure \ref{fig:overlap}), but those overlapping issues are assigned \textit{Critical} severity by SonarQube. We interpret this as evidence that issues flagged by both tools warrant high priority for refactoring operations. Since all overlapping issues between CodeScene and SonarQube are related to \textit{Code Complexity}, it indicates that \textit{Code Complexity} is one of the major design issues within AI IDE-generated projects.

\subsubsection{Relationship between Code Complexity Issues Identified by CodeScene.} Design issues identified by CodeScene include five categories related to code complexity: \textit{Overall Code Complexity}, \textit{Complex Method}, \textit{Complex Conditional}, \textit{Bumpy Road Ahead}, and \textit{Deep Nested Complexity}. Even though these issues are highly related, they represent different design issues for two reasons: (1) these issues are detected at different levels of granularity. For instance, \textit{Overall Code Complexity} is detected at the file level, while \textit{Complex Conditional} is identified at the statement level (e.g., an \texttt{if} statement). (2) Most of these design issue categories are detected by CodeScene using different metrics and indicate different levels of severity. For instance, \textit{Complex Method} design issues are detected based on the \textit{Cyclomatic Complexity} metric, while \textit{Complex Conditional} design issues are detected using the number of logical operations in conditional statements. Similarly, \textit{Deep Nested Complexity} is identified using the levels of nesting in the code.

 
\subsection{Implications}

\subsubsection{Implications for Practitioners}
Based on our results, we offer a set of recommendations for practitioners on how to use AI IDEs for their projects effectively. This includes recommendations focused on design quality and systematically utilizing AI IDEs for generating large projects.

\textit{Focus on design quality alongside functional correctness when adopting AI IDEs.} Practitioners should not only prioritize functional correctness (i.e., ``it works'') but also focus on design quality as a crucial dimension when adopting AI IDEs at scale. Practitioners should continuously monitor design quality metrics (e.g., complexity, duplication) using static analysis tools to keep the measurement within suggested ranges, thereby preventing the accumulation of design issues when generating projects using AI IDEs. 


\textit{Explicitly specify accessibility requirements for frontend code and use automated tools for checking accessibility.} Our results show that AI IDE-generated frontend UI implementations contain a considerable number of accessibility issues (e.g., missing ARIA labels, inadequate keyboard navigation, and missing form labels), which limit accessibility on devices that rely on alternative inputs such as keyboards, touch, or assistive technologies like screen readers. We recommend that practitioners state accessibility requirements (e.g., form controls should have associated labels, interactive elements should be keyboard-accessible, ARIA roles and labels should be assigned to UI elements) explicitly in specifications and task lists when using AI IDEs to generate projects, and use automated accessibility checks as part of the evaluation of AI IDE-generated frontends.


\textit{Invest more human effort at the high-level and delegate low-level tasks to AI IDEs.} When using AI IDEs for project generation, practitioners should allocate more effort to the project initialization, and requirements and design phases, particularly for the manual review and refinement of \texttt{requirements.md} and \texttt{tasklist.md}. We suggest a feature-driven strategy \cite{palmer2001practical} rather than attempting end-to-end generation in a single step for large-scale projects. Decomposing the project into independently testable features keeps each generation step within a manageable scope for the AI IDE and establishes clear checkpoints for human review. We suggest defining explicit acceptance criteria and delegating low-level implementation tasks to AI IDEs while retaining control over code quality.

\textit{Complete backend and database tasks first, and validate RESTful APIs before frontend integration.} In the context of AI IDE-generated projects, separating backend and frontend code generation by first completing backend and database tasks (i.e., as in our FD-HITL framework) enables early validation of RESTful APIs and localizes defects in business logic, data models, and API contracts before frontend code is generated. This isolates and resolves backend implementation and logical errors earlier, reducing the risk that those defects propagate into frontend issues or are misinterpreted as frontend issues.

\textit{Validate each feature with focused black-box testing before moving to the next feature to catch defects early.} Each generated feature should be followed by manual black-box testing of that feature alone, rather than relying only on system-level validation. Feature-wise validation cycles help identify implementation and logical errors early and enable follow-up prompts (e.g., bug-fix, logical-issue, or enhancement prompts). This practice reduces the accumulation of hidden defects that typically surface in later stages and are costly to fix.

\subsubsection{Implications for Researchers}
We present several implications for the future of AI-assisted software development below.

\textit{LLM-assisted code analysis for more critical design quality evaluation.} The code (static) analysis in our study revealed many issues, but it also required substantial manual filtering of false positives and was better suited to lower-level and file-level design issues than to system-wide architecture. Design issues such as inappropriate coupling or unclear module boundaries may therefore go undetected. Future work could explore complementary methods (e.g., LLM-assisted code analysis) with static analysis to assess the design quality of AI IDE-generated projects beyond what current static analysis tools can report.




\textit{Extend the FD-HITL framework to overcome critical design issues and improve frontend development.} First, our FD-HITL framework was effective in achieving functional correctness and project scale. Researchers could further extend the framework, for instance, by incorporating explicit design constraints (e.g., maximum method length, cyclomatic complexity limits, duplication thresholds) or by introducing a dedicated design-review phase after the main feature set is complete (e.g., refactoring to address identified design issues). Second, our project generation process relies on textual prompts to enable AI IDEs to build features. We did not employ other input modalities (e.g., visual Figma UI mockups), particularly to guide frontend design. Researchers could further extend the framework by incorporating multiple input modalities to better support development, especially on the frontend side.


\textit{Study the behavior of developers to inform the design of AI IDEs.} The behavior of developers utilizing AI IDEs may have a significant influence on both the success of the project under development and the quality of the generated project. Future research can conduct user studies to analyze which developer behavior patterns may limit the potential of AI IDEs. For instance, unlike the previous generation of code assistance tools, which provided simple snippet-level code suggestions that were easy to review and accept or reject, AI IDEs may generate a comparatively large amount of code, sometimes across multiple files. This may also lead developers to not thoroughly review the AI-generated code, causing them to over-rely on AI IDEs. Understanding these behavioral patterns can inform the design of AI IDE interfaces and features that better support developers in reviewing and generating code.

\section{Threats to Validity} \label{sec:validity}

\subsection{Internal Validity}
\textit{Subjectivity in manual evaluation.} We calculated the functional correctness of the generated projects using manual evaluation, which can introduce subjectivity bias. We mitigated this threat in two ways. First, two authors evaluated each project against the specifications in the \texttt{requirements.md} file and then reconciled their judgments to reach a consensus. Second, we documented our evaluation by capturing screenshots of all ``\textit{Complete}'' functional requirements, as described in Section~\ref{evaluation_process}. These screenshots are available in our replication package~\cite{replicationPackage}, making the evaluation verifiable.

\textit{Researcher and process influence.} The outcomes of project generation may depend on the experience and behavior of the researcher operating Cursor. Factors such as how prompts are phrased and how the feedback loop is provided can influence the resulting projects and the design issues they exhibit. Practical use of such tools typically requires human intervention and iterative refinement. To limit the impact of the ad hoc approach, we proposed and adopted a structured Feature-Driven Human-In-The-Loop (FD-HITL) framework that guides developers from high-level project description through to implementation (see Section~\ref{subsec:framework}). We followed this framework consistently across the generation of all 10 projects, ensuring a repeatable process rather than unstructured prompting.

\subsection{Construct Validity}
\textit{Limitations of static analysis tools.} CodeScene and SonarQube rely on rule-based static analysis and may miss certain design issues or report false positives that do not reflect genuine design issues in the given context. We addressed this threat by using two tools (i.e., CodeScene and SonarQube) in parallel to obtain complementary views on design issues. We also manually verified the reported issues by examining issue descriptions and confirming them in the source code. We classified an issue as a false positive only when it clearly contradicted the documentation of the project's underlying technology (see Section \ref{subsubsec:sonarmethod}). As a result of this manual verification, we identified and removed 1,612 false-positive issues (see Section~\ref{subsec:staticanalysis}).

\textit{Influence of project descriptions.} The content and structure of the curated project descriptions may affect the complexity, scope, and quality of the generated projects, and thus the set of design issues we observed. To reduce this threat, we had the first author draft the initial project descriptions, and then the other two co-authors reviewed them and provided feedback. We finalized the descriptions only after three authors agreed that the requirements were clear and that the descriptions were aligned across the study (see Section~\ref{subsec:formalgeneration}).

\textit{Reliability of qualitative categorization.} The design issue categories resulted from thematic analysis with one primary coder (the first author) and iterative review by the other two co-authors. Different coders might emphasize or group issues differently. To assess the reliability of our categorization, we calculated pairwise Cohen's kappa between the primary coder and the other reviewers. The resulting value of 0.88 indicates substantial agreement, supporting the consistency of our reported categories. We also used a negotiated agreement approach~\cite{campbell2013coding} to resolve remaining disagreements and had multiple co-authors review and provide feedback on the categories, as described in Section~\ref{subsec:dataanalysis}.

\subsection{External Validity}
\textit{Representativeness of the project sample.} Our findings are based on 10 projects generated from a fixed set of curated project descriptions. This sample may not capture the full diversity of real-world systems in terms of domain, scale, or development context. We sought to improve representativeness by designing project descriptions that span three application domains: mobile applications, Web applications, and utility tools. We also applied explicit inclusion and exclusion criteria (see Section \ref{subsec:formalgeneration}), so that each included project meets a minimum scale (e.g., at least 8K LoC) and has structural complexity similar to that of industrial systems. 

\textit{Technology and language specificity.} Design issues and the behavior of static analysis tools can vary across programming languages and technology stacks, which may limit the generalizability of our findings beyond the technologies we used. We mitigated this threat by generating projects with a variety of popular technologies (e.g., React, Spring Boot, Django, React Native) and programming languages (Java, JavaScript, PHP, Python), covering both frontend and backend contexts. This diversity allowed us to observe a broader range of design issues and report technology-specific issue patterns. We acknowledge that other technology stacks may exhibit different or additional design issues not observed in our study. 

\subsection{Reliability}
The reliability of empirical studies refers to the extent to which results and findings can be replicated under similar conditions. A potential threat to reliability arises from project generation, manual evaluation of the generated projects, and qualitative analysis of the identified design issues. To mitigate this threat, we followed a rigorous process for project generation, manual evaluation, and thematic analysis of issues (see Sections \ref{subsec:framework}, \ref{evaluation_process},  and \ref{subsec:dataanalysis}). We also made the dataset and replication materials available so that other researchers can replicate or extend our analysis~\cite{replicationPackage}.

\section{Conclusions and Future Work} \label{sec:conclusion}
This study explored whether AI IDEs with agentic capabilities can produce end-to-end large-scale software systems and what design issues those systems exhibit. We focused on Cursor as a representative AI IDE and conducted an empirical study in which we generated 10 large-scale projects from curated project descriptions using a Feature-Driven Human-In-The-Loop (FD-HITL) framework. We then assessed their design quality with two static analysis tools, CodeScene and SonarQube. Our findings show that Cursor can generate functional large-scale projects when used with the FD-HITL framework. The generated projects nevertheless contain a substantial number of design issues. We identified 1,305 issues in 9 categories by CodeScene and 3,193 issues in 11 categories by SonarQube. The most prevalent design issues include \textit{Code Duplication}, \textit{Code Complexity}, \textit{Long Methods}, \textit{Framework Best-Practice Violations}, \textit{Exception-Handling Issues}, and \textit{Accessibility Issues}. These issues frequently violate established design principles such as the \textit{Single Responsibility Principle (SRP)}, \textit{Separation of Concerns (SoC)}, and \textit{Don’t Repeat Yourself (DRY)}, and may pose long-term risks to maintainability and evolvability. We conclude that AI IDEs require continuous human judgment and, as currently stands, may not immediately replace developers' engineering skills, especially for high-level tasks (e.g., requirements engineering, architecture and design related tasks). Consequently, large-scale Cursor-generated projects warrant close review by experienced developers. 

Several directions for future work remain: (1) Replicating this study with other AI IDEs (e.g., Claude Code and Codex) and across different scales or application domains would strengthen the generalizability of our findings and reveal whether design issues are tool-specific or common across agentic code generation systems. (2) Developing tool support to detect or prevent the most frequent design issues (e.g., code duplication and code length) during or immediately after generation could help practitioners adopt AI IDEs more safely at scale. (3) Refining the FD-HITL framework or prompt strategies to reduce the occurrence of specific design issues (e.g., \textit{Code Duplication} and \textit{Code Complexity}) is a promising avenue for both research and practice. (4) Our dataset of Cursor-generated projects can be used to assess design quality using other code analysis tools and to identify additional design issues in these projects. LLM-powered code analysis tools can also be developed and explored to detect design issues in such projects.

\section*{Data Availability}
The dataset used in this work is available online at \cite{replicationPackage}.

\begin{acks}
This work has been partially supported by the National Natural Science Foundation of China (NSFC) with Grant No. 92582203 and 62402348 and the Major Science and Technology Project of Hubei Province under Grant No. 2024BAA008. 
\end{acks}

\end{sloppypar}

\bibliographystyle{ACM-Reference-Format}
\bibliography{basebib}










\end{document}